\documentclass[12pt]{article}
\usepackage[margin=0.8in]{geometry}
\usepackage[utf8]{inputenc}

\usepackage{amsmath,amssymb,amsfonts,amsthm}

\usepackage{breqn}

\usepackage{bold-extra}

\usepackage{graphicx}
\usepackage{xcolor}
\usepackage{hyperref}
\hypersetup{
  colorlinks   = true, 
  urlcolor     = blue, 
  linkcolor    = black, 
  citecolor   = black 
}
\usepackage[numbers,sort&compress]{natbib}
\let\OLDthebibliography\thebibliography
\renewcommand\thebibliography[1]{
  \OLDthebibliography{#1}
  \setlength{\parskip}{4pt}
  \setlength{\itemsep}{0pt plus 0.3ex}
}

\allowdisplaybreaks
\usepackage{tabularx}
\usepackage{bbm}
\usepackage{listings}
\usepackage{longtable}
\usepackage{array}

\usepackage{multicol}
\usepackage{enumitem}

\newcommand\refeq[1]{Eq.~(\ref{#1})}

\newcommand\refse[1]{Sect.~\ref{#1}}

\newcommand\citere[1]{Ref.~\cite{#1}}
\newcommand\citeres[1]{Refs.~\cite{#1}}

\def\reffi#1{\mbox{Fig.~\ref{#1}}}

\newcommand\refapp[1]{App.~\ref{#1}}

\newcommand{\htb}[1]{{\color{black} #1}}

\newcommand{\htm}[1]{{\color{black} #1}}

\newcommand{\htk}[1]{{\color{black} #1}}

\usepackage{soul}

\usepackage{blindtext}
\usepackage{floatrow}
\newfloatcommand{capbtabbox}{table}[][\FBwidth]
\usepackage[T1]{fontenc}
\usepackage[font=small,tableposition=top]{caption}
\DeclareCaptionLabelFormat{andtable}{#1~#2  \&  \tablename~\thetable}

\newcommand{\gev}{\, \mathrm{GeV}}
\newcommand{\tev}{\, \mathrm{TeV}}

\newcommand{\GW}[1]{{\color{black}#1}}

\graphicspath{{Plots/}}

\begin{document}

\thispagestyle{empty}

\def\thefootnote{\fnsymbol{footnote}}

\begin{flushright}
  DESY-22-127\\
  IFT--UAM/CSIC--22-015
\end{flushright}

\vspace*{1cm}

\begin{center}

  \textbf{{  \Large The trap in the early Universe: impact on the interplay \\[0.3em] between gravitational waves and LHC physics in the 2HDM}\\[1em]
                  }

  \vspace{0.5cm}

\setcounter{footnote}{3}

Thomas~Biekötter$^{1}$\footnote{thomas.biekoetter@desy.de},
Sven~Heinemeyer$^{2}$\footnote{Sven.Heinemeyer@cern.ch},
Jos\'e~Miguel~No$^{2,3}$\footnote{josemiguel.no@uam.es},\\[0.4em]
Mar\'ia~Olalla~Olea-Romacho$^{1}$\footnote{maria.olalla.olea.romacho@desy.de}
and~Georg~Weiglein$^{1,4}$\footnote{georg.weiglein@desy.de}

\vspace{0.5cm}

\textsl{$^{1}$Deutsches Elektronen-Synchrotron DESY, Notkestr.
85, 22607 Hamburg, Germany}

\vspace{0.1cm}

\textsl{$^{2}$Instituto de Física Teórica UAM-CSIC, Cantoblanco,
28049, Madrid, Spain}

\vspace{0.1cm}

\textsl{$^{3}$Departamento de F{í}sica Te{ó}rica, Universidad
Aut{ó}noma de Madrid (UAM), }\\
\textsl{ Campus de Cantoblanco, 28049 Madrid, Spain}

\vspace{0.1cm}

\textsl{$^{4}$II.\  Institut f\"ur  Theoretische  Physik, 
Universit\"at  Hamburg, Luruper Chaussee 149, }\\
\textsl{22761 Hamburg, Germany}

\vspace{0.1cm}

\begin{abstract}
We analyze the thermal history of the 2HDM and 
determine the parameter regions featuring a first-order electroweak  phase transition (FOEWPT) and also much less studied phenomena like high-temperature electroweak (EW) symmetry non-restoration and the
possibility of vacuum trapping (i.e.\ the Universe remains trapped in an EW-symmetric vacuum throughout the cosmological evolution, despite at $T=0$ the EW breaking vacuum is deeper).
We show that the presence of vacuum trapping impedes a first-order EW phase transition in 2HDM parameter-space regions previously considered suitable for the realization of 
electroweak baryogenesis.
Focusing then on the regions that do feature such a first-order transition, we show that the 2HDM parameter space that would yield a stochastic gravitational wave signal potentially detectable by the future LISA observatory is very contrived, and will be well probed by direct searches of 2HDM Higgs bosons at the HL-LHC, and (possibly) also via measurements of the 
self-coupling
of the Higgs boson at 125~GeV. This has an important impact on the interplay between LISA and the LHC regarding the exploration of first-order phase transition scenarios in the 2HDM: the absence of new physics indications at the HL-LHC would severely limit the prospects of a detection by LISA.
Finally, we demonstrate that 
as a consequence of the predicted enhancement of the self-coupling of the Higgs boson at 125~GeV
the ILC would be able to probe the majority of the 2HDM parameter space yielding a FOEWPT through measurements of the 
self-coupling, with a large improvement in precision with respect to the HL-LHC.

\end{abstract}

\end{center}

\renewcommand{\thefootnote}{\arabic{footnote}}
\setcounter{footnote}{0} 

\newpage

\tableofcontents

\section{Introduction}
\label{secintro}
The discovery of a \GW{Higgs boson at about 125 GeV} at the Large Hadron Collider (LHC)~\cite{Aad:2012tfa,Chatrchyan:2012ufa} was a milestone in our understanding of the laws of nature. At present, the properties of the 
\GW{detected} Higgs boson agree well with the predictions of the Standard Model (SM)~\cite{CMS:2022dwd,ATLAS-CONF-2022-050}, \GW{but at} the current $\sim 10\%$ precision \GW{of the} Higgs \htm{boson} signal strength measurements at the LHC 
\GW{the experimental results are also in agreement with}
a plethora of extensions of the SM. Such extensions could address the various shortcomings of the SM. In particular, 
the ingredients of the SM are not sufficient to generate the observed matter-antimatter asymmetry of the Universe, the so-called \textit{baryon asymmetry of the Universe} (BAU)~\cite{Huet:1994jb,Gavela:1994ds,Gavela:1994dt,Kajantie:1996mn}.

It is well-known that the possible existence of extra Higgs doublets beyond the SM~\cite{Lee:1973iz,Kim:1979if,Wilczek:1977pj} could allow for the generation of 
the BAU via electroweak (EW) 
baryogenesis~\cite{Cline:1996mga,Fromme:2006cm,Cline:2011mm,Dorsch:2016nrg}
(see~\cite{Cohen:1993nk,Trodden:1998ym,Morrissey:2012db} for general reviews on EW baryogenesis). Such a scenario requires a (strong) first order EW phase transition
(FOEWPT), which provides the required out-of-equilibrium conditions for baryogenesis in the early Universe~\cite{Sakharov:1967dj}. Scenarios featuring a FOEWPT also have the remarkable feature that they would lead to a stochastic gravitational wave (GW) background that could potentially be detectable with future space-based gravitational wave 
observatories like the Laser Interferometer Space Antenna (LISA)~\cite{Caprini:2019egz,Auclair:2022lcg}.

Theories with extended (non-minimal) Higgs sectors may also feature a rich variety of possible cosmological histories of the EW vacuum, as compared to the SM.
In the SM the spontaneously broken EW symmetry is restored at high temperatures.
The EW gauge symmetry is broken dynamically when the Universe cools down, which in the SM happens (for the measured Higgs boson mass $m_h \approx 125$ GeV) via a smooth cross-over transition~\cite{Kajantie:1996mn}.
It was however established long ago~\cite{Weinberg:1974hy} that theories with additional scalar fields can give rise to different symmetry-breaking patterns as a function of the temperature of the Universe:~a symmetry might remain broken at all temperatures, only be restored in an intermediate temperature region, or even be broken \textit{only} at \GW{high} temperatures, being unbroken at zero temperature. All these alternatives, so-called ``symmetry non-restoration'' (SnR)  scenarios, may occur for non-minimal Higgs sectors, with a potential impact on the viability of the ``vanilla'' EW baryogenesis mechanism.
A further possibility in the thermal history of extended Higgs sectors
is vacuum trapping: at zero temperature the EW vacuum would exist as the deepest \GW{(or sufficiently long-lived)} vacuum of the potential, \GW{but} during the thermal evolution of the Universe the conditions for the on-set of the
EW phase transition would never be fulfilled. \GW{Thus,} the Universe would remain
trapped in a (higher-energetic) non-EW vacuum, yielding an unphysical scenario. \GW{Parameter regions giving rise to vacuum trapping can therefore be excluded~\cite{Cline:1999wi,
Baum:2020vfl,Biekotter:2021ysx}.}

\vspace{1mm}

In this work we analyze in detail the thermal history of the Universe in the two-Higgs-doublet model (2HDM) (see~\citere{Branco:2011iw} for a review).
\GW{In particular, we investigate}
the occurrence of a FOEWPT as needed for EW baryogenesis, as well as the production of GWs potentially observable by LISA. We show the important impact that SnR and vacuum-trapping phenomena (which can appear in the 2HDM despite its relatively simple structure) have in shaping the 2HDM regions of parameter space where baryogenesis and/or GW production are possible. In particular, we demonstrate that vacuum-trapping reduces the 2HDM parameter range for which a GW signal from a FOEWPT would be observable by LISA to a very fine-tuned parameter-space region.   

In addition, focusing on the type~II 2HDM, we investigate the connection between the thermal history of the early Universe, particularly regarding a possible FOEWPT, and phenomenological signatures at colliders 
(see~\citeres{Dorsch:2013wja,Dorsch:2014qja,Basler:2016obg,Dorsch:2017nza,Bernon:2017jgv,Goncalves:2021egx} for earlier analyses of this connection in the 2HDM): we study the new BSM Higgs boson signatures that are favored by scenarios with a FOEWPT. We demonstrate that ongoing and future LHC searches in final states with top quarks will probe the vast majority of the 2HDM parameter-space region yielding a strongly FOEWPT, already covering the entire region accessible via GW observations by LISA.
We also analyze the connection between a FOEWPT and a large enhancement of the \GW{self-coupling of the Higgs boson at} 125~GeV 
with respect to its SM value~\cite{Noble:2007kk,Huang:2015tdv}.
We show that \GW{experimental information on} the Higgs boson self-coupling at the High-Luminosity (HL)-LHC and particularly at the International Linear Collider (ILC) \GW{will provide} a very promising \GW{route for probing} FOEWPT scenarios in the 2HDM (and more broadly, in extended Higgs sectors).

Our work is organized as follows: In section~\ref{sec:2hdm} we discuss the Higgs sector of the 2HDM including higher-order corrections. The effects of the thermal (early universe) evolution of the EW vacuum are reviewed in section~\ref{sec:ThermalHist}, where we also discuss the 
general aspects of SnR, vacuum trapping, and GW production during a FOEWPT.
Our analysis of the cosmological evolution of the scalar vacuum in different regions of the 2HDM parameter space is then presented in section~\ref{secnumanal}, and the connection with both GW production and collider phenomenology is discussed, providing a critical view on the interplay between these two. 
Our conclusions can be found in section~\ref{conclu}.

\newpage
 
\section{The \GW{Higgs sector of the} 2HDM} 
\label{sec:2hdm}
In this section we discuss the aspects of the 2HDM \GW{that are relevant 
for 
this} work: we first briefly review the CP-conserving 2HDM with a softly broken $\mathbb{Z}_{2}$ symmetry
(see e.g.~\citere{Branco:2011iw} for further details) and specify our notation and conventions; we then discuss the theoretical and experimental
constraints that shape the allowed
parameter space of the model, before describing the form of the one-loop effective potential and the renormalization group running of 2HDM scalar couplings.

\subsection{Model definition and notation}
\label{sec:treelevel2hdm}
The tree-level potential of the CP-conserving 2HDM with a softly-broken $\mathbb{Z}_{2}$ symmetry is given by
\begin{align}
V_{\text{tree}}&=m_{11}^{2}\left|\Phi_{1}\right|^{2}+m_{22}^{2}\left|\Phi_{2}\right|^{2}-m_{12}^{2}\left(\Phi_{1}^{\dagger}\Phi_{2}+\text{h.c.}\right)+\frac{\lambda_{1}}{2}\left(\Phi_{1}^{\dagger}\Phi_{1}\right)^{2}+\frac{\lambda_{2}}{2}\left(\Phi_{2}^{\dagger}\Phi_{2}\right)^{2} \notag \\
&+\lambda_{3}\left(\Phi_{1}^{\dagger}\Phi_{1}\right)\left(\Phi_{2}^{\dagger}\Phi_{2}\right)+\lambda_{4}\left(\Phi_{1}^{\dagger}\Phi_{2}\right)\left(\Phi_{2}^{\dagger}\Phi_{1}\right)+\frac{\lambda_{5}}{2}\left[\left(\Phi_{1}^{\dagger}\Phi_{2}\right)^{2}+
\mathrm{h.c.}\right],
\label{Tree-Level-Potential}
\end{align}
where all the parameters are 
real as a result of hermiticity and
CP-conservation.
The $\mathbb{Z}_{2}$ symmetry in~\eqref{Tree-Level-Potential},
$\Phi_{1}\rightarrow\Phi_{1},\,\Phi_{2}\rightarrow-\Phi_{2}$ is broken softly by the
$m_{12}^{2}$ term.
We expand the fields around the EW minimum as follows,
\begin{equation}
\Phi_1 =
\begin{pmatrix}
\phi_1^+ \\
\left(v_1 +  \rho_1 + \mathrm{i} \sigma_1 \right)
    / \sqrt{2}
\end{pmatrix} \ , \quad
\Phi_2 =
\begin{pmatrix}
\phi_2^+ \\
\left(v_2 +   \rho_2 + \mathrm{i} \sigma_2 \right)
    / \sqrt{2}
\end{pmatrix}, \quad
\end{equation}
where $v_{1}$ and $v_{2}$ are the field vevs for the two Higgs doublets (at zero temperature), \GW{and} the EW scale is defined as $v=\sqrt{v_{1}^{2}+v_{2}^{2}}\approx 246 \gev$. After spontaneous symmetry breaking, the CP-conserving 2HDM gives rise to five physical mass eigenstates in the scalar sector: two CP-even neutral scalars $h$ and $H$, one CP-odd neutral pseudoscalar $A$ and a pair of charged states $H^{\pm}$. In addition, there are one neutral and two charged
massless Goldstone bosons, $G^{0}$ and $G^{\pm}$ respectively, which are absorbed into
longitudinal polarisations of the gauge bosons $Z$ and $W^{\pm}$, respectively.

For the CP-odd and the charged scalar sectors, the mass eigenstates are related to the gauge eigenstates by an orthogonal rotation by the angle $\beta$, defined as $t_{\beta} \equiv \tan \beta = v_{2}/v_{1}$.
For the CP-even sector, the mixing angle $\alpha$ can be treated as a free parameter. The parameters $\alpha$ and $\beta$ control the coupling strength of the scalar particles to fermions and gauge bosons (see e.g. \citere{Branco:2011iw} for the explicit
form of the effective couplings). Therefore, it is convenient to perform 
our analysis in terms of the particle masses of the Higgs sector and the mixing angles 
since 
\GW{their phenomenological impact is transparent}.
We choose the following set of independent parameters:
\begin{equation}
t_{\beta} \ , \; m_{12}^{2} \ ,
\; v \ , \; \cos (\beta - \alpha) \ ,
\; m_{h} \ , \;m_{H} \ , \; m_{A} \ ,
\; m_{H^{\pm}}\ .
\label{input_parameters}
\end{equation}
Here, $m_h$ and $m_H$ are the masses of the CP-even Higgs bosons, $m_A$ is the
mass of the CP-odd Higgs boson, and $m_{H^\pm}$ is the mass of the
charged Higgs bosons, respectively. The parameter $\cos(\alpha - \beta)$
is chosen in order to have a measure as to how closely the state $h$,
which in the following plays the role of the discovered Higgs boson
at \GW{$m_h \approx 125\gev$},
resembles the properties of a SM Higgs boson.
In the so-called alignment limit~\cite{Gunion:2002zf} $\cos(\alpha - \beta) = 0$, the \GW{lowest-order} couplings of $h$ to the SM particles are precisely as
predicted by the SM, whereas $\cos(\alpha - \beta) \neq 0$ gives rise
to deviations of the couplings of $h$ from their SM values.
The relations between the set of input parameters shown in \refeq{input_parameters}
and the Lagrangian parameters shown in \refeq{Tree-Level-Potential} can
be found in \citere{Branco:2011iw}.

The $\mathbb{Z}_2$ symmetry in \refeq{Tree-Level-Potential} is extended to the
Yukawa sector of the theory in order to avoid tree-level flavour-changing neutral currents (FCNCs): as the two fields $\Phi_{1}$ and $\Phi_{2}$ transform differently 
under the $\mathbb{Z}_{2}$ symmetry, they cannot be coupled
both to the same SM fermions, leading to the absence of tree-level
FCNCs. There are four 2HDM configurations/types that avoid FCNCs at tree-level, characterized by the $\mathbb{Z}_{2}$ charge assignment of the fermion fields in the Yukawa sector. 
As a consequence, in addition to the values of the free parameters shown in
\refeq{input_parameters} the Yukawa type of the 2HDM has to be specified.
Here we will concentrate on the 2HDM of type~II, and focus on the alignment limit.

\vspace{1mm}

The parameter space of the CP-conserving 2HDM, 
which has \GW{eight} independent parameters \GW{as specified above}, is restricted by various experimental and theoretical constraints. To implement these in our analysis, we make use of several public codes. We scan the 2HDM parameter space with the code
\texttt{ScannerS} \cite{Coimbra:2013qq, Muhlleitner:2020wwk}
in terms of the set of parameters shown in
\refeq{input_parameters}. \texttt{ScannerS} 
checks whether the parameter point under investigation is in agreement with perturbative unitarity, boundedness from below and vacuum stability at zero temperature.
Concerning the experimental
constraints, \texttt{ScannerS} also ensures that a parameter point
is in agreement with bounds coming from flavour-physics observables~\cite{Haller:2018nnx} and electroweak precision
observables (EWPO)~\cite{Grimus:2007if,Grimus:2008nb,Haller:2018nnx}.\footnote{The check for the agreement with the EWPO (carried out on the basis of the oblique parameters) does
not take into account the new measurement of the $W$-boson mass reported recently
by the CDF collaboration~\cite{CDF:2022hxs}, which
is in significant tension with the SM predictions.} 
In addition, we make use of \texttt{HiggsSignals}~\cite{Bechtle:2013xfa,
Stal:2013hwa,Bechtle:2014ewa,Bechtle:2020uwn} and  
\texttt{HiggsBounds}~\cite{Bechtle:2008jh,Bechtle:2011sb,
Bechtle:2013gu,Bechtle:2013wla,Bechtle:2020pkv}
to incorporate bounds
from measurements of the properties of the experimentally detected Higgs boson \GW{at about $125 \gev$} and searches for additional scalar states, respectively.
The required cross sections and branching ratios
of the scalars have been obtained with the help of
\texttt{SusHi}~\cite{Harlander:2012pb}
and \texttt{N2HDECAY}~\cite{Engeln:2018mbg}, respectively.

\subsection{2HDM \GW{effective} potential and renormalization conditions}
\label{sec: renorm2hdm}

The zero-temperature effective potential (see e.g.~\citere{Quiros:1999jp} for a review) includes the effect of radiative corrections into the scalar potential of 
the theory. At one-loop, the effective potential $V_{\mathrm{eff}}$ for the 2HDM is given by 
\begin{equation}
\label{Veff_Zero}
V_{\mathrm{eff}} = V_{\mathrm{tree}} + V_{\rm CW}+ V_{\rm CT}\, ,
\end{equation}
where $V_{\rm tree}$ is the 2HDM tree-level potential given in
\refeq{Tree-Level-Potential}, $V_{\rm CW}$
represents the one-loop radiative corrections in the form of the Coleman-Weinberg
potential~\cite{Coleman:1973jx}, and $V_{\rm CT}$ contains UV-finite counterterm contributions that are specified below. $V_{\rm CW}$ is given in the $\overline{\mathrm{MS}}$ renormalization \GW{scheme} by
\begin{equation}
V_{\text{CW}}(\phi_i) = \sum_{j}\frac{n_{j}}{64\pi^{2}}(-1)^{2s_{i}} \,
m_{j}(\phi_i)^{4}
\left[\ln\left(\frac{|m_{j}(\phi_i)^{2}|}{\mu^{2}}\right)-c_{j}\right],
\label{CW_potential}
\end{equation}
where $m_{j}(\phi_i)$ are the field-dependent tree-level masses, $s_{j}$ the particle spins and $n_{j}$ the corresponding numbers of degrees of freedom. The constants $c_{j}$ arise from the $\overline{\mathrm{MS}}$ renormalization \GW{prescription},
with $c_j=5/6$ for gauge bosons and $c_{j}=3/2$ for scalars and fermions. We set the
renormalization scale $\mu$ to be equal to the SM EW vev, $\mu = v$.
In the 2HDM, the sum in~\refeq{CW_potential} runs over the neutral scalars $\Phi^{0}=\{h,\,H,\,A,\,G^{0}\}$, the charged scalars $\Phi^{\pm}=\{H^{\pm},\,G^{\pm}\}$, the longitudinal and transversal gauge bosons,
$V_{L}=\{Z_{L},\,W_{L}^{+},\,W_{L}^{-}\}$ and $V_{T}=\{Z_{T},\,W_{T}^{+},\,W_{T}^{-}\}$ and the SM quarks $q$ and
leptons $\ell$. The degrees of freedom $n_{j}$ for the species of each type are
\[
n_{\Phi^{0}}=1 \,, \quad n_{\Phi^{\pm}}=2\, ,\quad   n_{V_{T}}=2\, ,\quad
n_{V_{L}}=1\, , \quad n_{q}=12 \,,\quad n_{\ell}=4\, .
\]
The omission of ghost contributions is 
\GW{due to the choice of}
evaluating the Coleman-Weinberg potential
in the Landau gauge.\footnote{Discussions on the gauge dependence of the effective
potential in the context of the electroweak phase transition can
be found in \citeres{Dolan:1973qd, Patel:2011th, Wainwright:2011qy, Garny:2012cg}.}

\vspace{1mm}

In order to simplify our analysis, we require the zero-temperature loop-corrected vacuum expectation values, scalar masses and mixing angles to be equal to their tree-level values, and we refer to this prescription as ``on-shell'' (OS) renormalization in the remainder of the paper.
To achieve this,
we add a set of UV-finite counterterms $V_{\rm CT}$
to the effective potential, given by
\begin{equation}
V_{\rm CT}=\sum_{i}\frac{\partial V_{0}}{\partial p_{i}}
\delta p_{i} \ ,
\label{eq:introparacounters}
\end{equation}
where the $p_{i}$ denote the parameters of the tree-level potential.
In order to fulfill the conditions
mentioned above, $V_{\rm CT}$ is required to
satisfy the following renormalization conditions,
\begin{align}
\partial_{\phi_{i}}V_{\rm CT}(\phi)\left|_{\left\langle
\phi\right\rangle_{T=0}}\right.&=-\partial_{\phi_{i}}V_{\rm CW}(\phi)
\left|_{\left\langle \phi\right\rangle _{T=0}}\right. \ , \\
\partial_{\phi_{i}}\partial_{\phi_{j}}V_{\rm CT}(\phi)
\left|_{\left\langle \phi\right\rangle _{T=0}}\right.&=
-\partial_{\phi_{i}}\partial_{\phi_{j}}V_{\rm CW}(\phi)
\left|_{\left\langle \phi\right\rangle _{T=0}}\right. \ ,
\end{align}
where ${\left\langle \phi\right\rangle _{T=0}}$ corresponds to the tree-level
vacuum configuration at zero temperature.
To compute the finite set of counterterms,
we made use of the public code
\texttt{BSMPT}~\cite{Basler:2018cwe, Basler:2020nrq}.

The effective potential explicitly 
depends on the renormalization scale $\mu$.
As mentioned 
above, in our
numerical analysis this scale was set 
equal to $v$, which is the relevant energy
scale for the physics at the EW scale.
Although the OS prescription discussed above
gives rise to a partial cancellation of
the renormalization scale dependence, there
still remains a dependence on $\mu$ for
the quantities that describe the EW phase
transition, in particular once the thermal
effects are taken into account (see
\refse{sec:ThermalHist}).
For a single parameter point,
the residual scale dependence
is relevant for the prediction of parameters
like the transition strengths or the SNR
of the associated GW signal,
where the predicted values can
vary substantially for different choices
of $\mu$~\cite{Croon:2020cgk}.
However, the logarithmic $\mu$-dependence 
(see \refeq{CW_potential})
is much milder compared to the
dependence on the
2HDM model parameters (such as
the masses of the BSM Higgs bosons).
As a result, changes in the predictions
for the EW phase transition
arising from a modification of $\mu$ can be compensated
 by very small changes of the
model parameters.
Given the fact that BSM theories like
the 2HDM have several free parameters,
in our phenomenological analysis we are mainly
interested in 
identifying regions
of the 2HDM parameter space that are suitable for
the realization of EW SnR, a strong FOEWPT and
observable GW signals.
As will be discussed in the numerical analysis,
the shapes and sizes of these regions
are 
very sensitive to a variation
of the 2HDM model parameters. 
Since the renormalization scale
dependence is much milder, the 
qualitative distinction between the different parameter regions
is only marginally
affected by variations of $\mu$.
Therefore, also our conclusions about the
interplay between the possibility of
a detection of GWs and the physics at
the LHC in the context of the 2HDM
are not significantly
affected by the renormalization-scale
dependence.
In order to illustrate this point in more detail,
we supplement our numerical discussion with
a comparison of the $\mu$-dependence with
the dependence on the model parameters
for an example scenario in \refapp{app:scale}.

\subsection{Scale dependence and perturbativity of scalar couplings}
\label{sec:rges}

The renormalization group evolution of the quartic scalar couplings $\lambda_i$ can provide \GW{significant} constraints on the viable region of the 2HDM parameter space.
Even if $\lambda_i(\mu)$ are perturbative at an energy scale $\mu = v$, the running of the parameters may drive the 2HDM quartic couplings into a non-perturbative regime. Depending on the values of $\lambda_i(v)$ this can happen already at relatively low energy scales (i.e.~not far from the EW scale). 
Hence, as 
\GW{an important ingredient of} our analysis, 
we solve the renormalization group equations (RGEs) for the model parameter points discussed, and require that the
$\lambda_i(\mu)$ remain 
below the perturbativity bound $4 \pi$ for any value of the energy scale $\mu$ up to the physical scalar masses of the theory $m_j$, i.e.~$\lambda_i(\mu) < 4\pi$ for $\mu \le m_j$ ($ \forall j$). This provides a (minimal) theoretical consistency condition on the 2HDM parameter space in relation to renormalization group evolution.

We have numerically solved the RGEs taking into account the one-loop and two-loop contributions to the $\beta$-functions of the model parameters computed with the help of
the public code \texttt{2HDME} \cite{Oredsson:2018vio}. 
In order to obtain $\overline{\rm MS}$ parameters $p^{\overline{\rm MS}}$ (as required by \texttt{2HDME}) from our OS parameters $p^{\rm OS}$, we transformed
the parameters according to
\begin{align}
p^{\rm OS}(\mu_0) + \delta p^{\rm OS}(\mu_0) &=
p^{\overline{\rm MS}}(\mu_0)
+ \delta p^{\overline{\rm MS}}(\mu_0) \\
\Rightarrow \ p^{\overline{\rm MS}}(\mu_0) &=
p^{\rm OS}(\mu_0) +
\delta p(\mu_0) \ ,
\end{align}
where $\delta p^{\overline{\rm MS}}$
are the parameter counterterms in the
$\overline{\rm MS}$ scheme \GW{(consisting only of a UV-divergent contribution)}, and
$\delta p^{\rm OS} =\delta p^{\overline{\rm MS}}
+ \delta p$ are the
OS counterterms that additionally contain
the UV-finite shifts $\delta p$ introduced in
\refeq{eq:introparacounters}.
We also stress that thermal effects, to be discussed in the next section, introduce the temperature of the system $T$ as a relevant energy scale. Then, 
for the study of the scalar potential at temperatures substantially
larger than the EW scale, $T \gg v$ (targeted towards the 
\GW{investigation}
of whether the EW symmetry is restored 
\GW{for high temperatures}, see section~\ref{sec:SnR}), 
we must also require $\lambda_i(\mu = T)$ to be perturbative.

\section{Thermal effects and thermal evolution}
\label{sec:ThermalHist}
We now describe the addition of thermal corrections to the effective potential $V_{\mathrm{eff}}$, which allows \GW{the extension of} the analysis of vacuum configurations to finite temperatures. Subsequently, we discuss several phenomena which may occur in the thermal evolution of the vacuum configuration of a (multi-) Higgs potential: a first-order EW phase transition, possibly with an accompanying stochastic signal of gravitational waves (GW); the non-restoration of EW symmetry at high temperatures (see \cite{Biekotter:2021ysx,Carena:2021onl}); and the trapping of the vacuum in an unbroken EW configuration (see \cite{Biekotter:2021ysx,Baum:2020vfl}). In this section we give general details of our computational set-up for these phenomena, and analyze their impact on the 2HDM parameter space in section~\ref{secnumanal}.

\subsection{Finite-temperature effective potential}
\label{sec:2hdmfiniteT}

\label{finiteTeffpot}
At finite temperature $T$, the one-loop effective potential from Eq.~\eqref{Veff_Zero} receives thermal corrections $V_T$, given by~\cite{Dolan:1973qd,Quiros:1999jp}
\begin{equation}
V_{T} (\phi_i)=\sum_{j} \, \frac{n_{j} \, T^{4}}{2\pi^{2}}\,
J_{\pm}\left(\frac{m_{j}^2(\phi_i)}{T^{2}}\right)  ,
\label{thermal_potential}
\end{equation}
where the thermal integrals $J_{-}$ for bosons and $J_{+}$ for fermions, 
\GW{respectively,}
are defined by
\begin{equation}
J_{\pm}\left(\frac{m_{j}^2(\phi_i)}{T^{2}}\right) = \mp
\int_{0}^{\infty}dx\,x^{2}\,\log\left[1\pm\exp\left(-\sqrt{x^{2}+\frac{
m_{j}^2(\phi_i)  }{T^{2}}}\right)\right]\, .
\label{thermalfunctions}
\end{equation}
Besides the degrees of freedom considered in~\refeq{CW_potential}, the sum in~\refeq{thermal_potential} includes the photon,
which acquires an effective thermal mass at finite temperature (and therefore must be
included in the sum \GW{of} \refeq{thermal_potential}
in spite of being massless at
${T=0}$).

In addition, the breakdown of the
fixed-order result of the
perturbative expansion
in the high-temperature limit 
is a well-known problem of finite-temperature field theory.
It can be addressed through the resummation of a certain
set of higher-loop diagrams~\cite{Gross:1980br,Parwani:1991gq,Arnold:1992rz},
the so-called daisy contributions (see~\citere{Quiros:1999jp} for a review).
There are various daisy-resummation prescriptions in the literature.\footnote{See \cite{Biekotter:2021ysx} for a comparison between the Arnold-Espinosa and the
Parwani resummation methods.
Recent computations of the characteristics
of first-order phase transitions that go
beyond the usual daisy-resummed approach
have been performed, for instance, in
\citeres{Croon:2020cgk,Schicho:2022wty},
where it was shown that
two-loop contributions to the effective
potential can be sizable. We leave a
discussion of the 2HDM potential beyond
the one-loop level for future studies.}
We here follow the Arnold-Espinosa
method~\cite{Arnold:1992rz}, which amounts to
resumming the infrared-divergent contributions from
the bosonic Matsubara zero-modes by adding
the additional contribution $V_{\text{daisy}}$ to
the one-loop effective potential at
finite temperature. $V_{\text{daisy}}$ is given by
\begin{equation}
V_{\text{daisy}}=-\sum_{k}\frac{T}{12\pi}
\left( \left(\bar{m}_k^2(\phi)\right)^{\frac{3}{2}} - \left( m_{k}^2(\phi)\right) ^{\frac{3}{2}} \right) \,  ,
\label{Daisy1resummation}
\end{equation} 
where the sum in $k$ runs over the bosonic degrees of freedom yielding
infrared-divergent contributions, and 
$\bar{m}_k^2$ denotes their corresponding squared thermal
masses~\cite{Carrington:1991hz}, which have been obtained
as in \citere{Basler:2016obg}.
In the 2HDM, the sum in $k$ runs 
over all the fields $\phi \in \left\{ \Phi^{0},\Phi^{\pm}, V_{L}, \gamma_L \right\}$, where $\gamma_L$ is the longitudinal polarisation of the photon, which acquires a 
mass at finite temperature.

\vspace{2mm}

With the inclusion of thermal corrections, the 2HDM (finite-temperature) one-loop effective potential with daisy-resummation reads
\begin{equation}
    \label{Veff_Finite}
    V_{\mathrm{eff}} = V_{\mathrm{tree}} + V_{\rm CW} + V_{\rm CT} + V_{\rm T}
    + V_{\rm daisy} \ .
\end{equation}
with $V_{\mathrm{tree}}$, $V_{\rm CW}$ and $V_{\rm CT}$ given in section~\ref{sec:2hdm}, and $V_T$, $V_{\rm daisy}$ described above.

\subsection{Characterization of first-order phase transitions}
\label{sec:FOEWPT}

In this section we briefly review the analysis of the
thermal evolution of the effective potential $V_{\mathrm{eff}}$, for which we use the public
code~\texttt{CosmoTransitions}~\cite{Wainwright:2011kj}). We are particularly interested in the occurrence of a first-order EW phase transition (FOEWPT), which requires the coexistence (at some temperature in the early Universe) of the EW minimum and another minimum. At the critical temperature $T_c$ these two minima are degenerate, while for $T < T_c$, the EW minimum has lower energy than the other (metastable) vacuum.
In this case, the occurrence of the FOEWPT depends on the transition rate per unit time and unit volume for the phase transition
from the metastable or false vacuum into the true (EW)
vacuum~\cite{Coleman:1977py,Callan:1977pt,Linde:1980tt,Linde:1981zj},
\begin{equation}
\Gamma(T)=A(T)\,e^{-S_{3}(T)/T},
\label{eqfuncdeter}
\end{equation}
\GW{where $S_3$ denotes} the three-dimensional
action for the ``bounce'' (multi-)field
configuration $\phi_{\rm{B}}$ that interpolates between the metastable and EW
vacua for $T < T_c$, 
\begin{equation}
S_{3}(T) = 4\pi \int r^{2} dr \, \left[\frac{1}{2}
\left(\frac{d\phi_{\rm{B}}}{dr}\right)^{2}+
V_{\rm eff} \left(\phi_{\rm{B}},T\right)\right]\, .
\label{eq:bounceaction}
\end{equation}
Specifically, the bounce $\phi_{\rm{B}}$ is the configuration of scalar fields $\phi$
that solves the equations of motion derived from the action~\eqref{eq:bounceaction} with boundary conditions $\left. d\phi/dr\right|_{r=0}=0$ and approaching the false vacuum at $r \to \infty$. 
Physically, $\phi_{\rm{B}}$ describes a bubble of the true vacuum phase nucleating in the false vacuum background.
The prefactor $A(T)$ is a functional determinant~\cite{Callan:1977pt} given
approximately by $A(T) \sim T^4 \, (S_3/2\pi T)^{3/2}$~\cite{Linde:1980tt}.
The onset of the FOEWPT occurs when the time integral of the
transition rate~\eqref{eqfuncdeter} within a Hubble volume $H$ becomes of order one. This defines the nucleation temperature $T_{n}$ (see e.g.~\cite{Espinosa:2008kw}) as
\begin{equation}
\int_{T_n}^{T_c} \frac{T^4}{H^4} \frac{A(T)}{T} \,e^{-S_{3}(T)/T}\, dT \sim 1~,
\,
\label{eq:_TN}
\end{equation}
where we have used the time-temperature relation in an expanding Universe assumed to be dominated by radiation. The Hubble parameter $H$ is given by
$H^2 = (8\,\pi^3 g_{\text{eff}}(T) \, T^4)/(90\, \rm{M}_{\rm{Pl}}^2)$, with $g_{\text{eff}}(T)$ the effective number of relativistic degrees of freedom at a temperature $T$ and $\rm{M}_{\rm{Pl}}= 1.22 \times 10^{19} \gev$ the Planck mass.
\refeq{eq:_TN}~roughly yields~\cite{Auclair:2022lcg}
\begin{equation}
S_3(T_n) / T_n \sim 140 \, ,
\label{cond_trans}
\end{equation}
\GW{as} the requirement for the occurrence of a FOEWPT.
The possibility that the condition \eqref{eq:_TN}
is not satisfied for any temperature below
the critical temperature $T_c$ will be discussed in section~\ref{sec:vactrapp}.

\vspace{2mm}

On general grounds, a cosmological first-order phase transition can be characterized by four macroscopic parameters which we specify in the following. These parameters are obtained from the microscopic properties of the underlying particle physics model,
in our case the 2HDM. 
As will be discussed in more detail in section \ref{sec:gws}, these parameters also determine the predictions of the amplitude and the spectral shape of the stochastic GW background that is generated during the first-order phase transition.

The first key parameter is the temperature $T_*$ at which the phase transition
takes place. The second parameter, $\alpha$, measures the strength of the phase transition. Following \citeres{Caprini:2019egz,Auclair:2022lcg}, we define $\alpha$ as
the difference of the trace of the
energy-momentum tensor between the two (false and true vacua) phases, normalised to the radiation background energy density, i.e.\
\begin{equation}
    \alpha =
    \frac{1}{\rho_{R}}
    \left(
    \Delta V(T_{*})-\left.\left(
    \frac{T}{4}\frac{\partial
    \Delta V(T)}{\partial T}\right)\right|_{T_{*}}
    \right) \ .
\label{eqalpha}
\end{equation}
Here $\Delta V(T)=V_{f}-V_{t}$, with $V_f \equiv V_{\rm eff}(\phi_f)$ and $V_t \equiv V_{\rm eff}(\phi_t)$ being the values of the potential in the false and the true vacuum,
respectively.\footnote{In some studies (see, for instance, \citeres{Goncalves:2021egx,
Kanemura:2022ozv} for 2HDM analyses) the parameter $\alpha$ has been defined instead as the latent heat released during the transition divided by $\rho_R$, in which case the factor $1/4$ in the second term in \refeq{eqalpha} is absent.
However, recent studies have shown that the definition used here yields a better
description of the energy budget available for the production of GW waves compared to
a definition of $\alpha$ by means of the pressure difference or the energy
difference~\cite{Giese:2020rtr}.} $\rho_{R}$ is the background energy density assuming a radiation dominated Universe, i.e.\ $\rho_{R}=\pi^2 g_{\text{eff}}(T_{*})T_{*}^4/30$.
We also note that for cosmological phase transitions which are not very strong, i.e.~$\alpha \ll 1$, the transition temperature $T_*$ can be identified
with the nucleation temperature $T_n$ defined by \refeq{eq:_TN}~\cite{Caprini:2019egz}, since the temperature at the onset of the transition is approximately equal to the temperature for which true vacuum bubbles collide and the phase transition completes.

The third quantity is 
the inverse duration of the phase transition in Hubble units, $\beta/H$.
It can generally be expressed (see~\cite{Ellis:2018mja} for a discussion) in terms of the derivative of the action $S_{3}$ with respect to the temperature evaluated at the
time of the phase transition,
\begin{equation}
\frac{\beta}{H} =
T_{*}\left.\left(\frac{d}{dT}\frac{S_{3}(T)}{T}\right)\right|_{T_{*}}.
\label{beta}
\end{equation}

The fourth quantity that characterizes a cosmological first-order phase transition is the bubble wall velocity $v_{\rm w}$. So far, except for the case of ultrarelativistic bubbles~\cite{Bodeker:2009qy,Bodeker:2017cim}
the computation of $v_{w}$ is generally a very complicated task that requires solving a coupled system of Boltzmann and scalar field equations in a fairly model-dependent approach~(see~\citeres{Moore:1995si,Konstandin:2014zta,Kozaczuk:2015owa,Dorsch:2018pat,BarrosoMancha:2020fay,Laurent:2020gpg,Dorsch:2021nje,Ai:2021kak,Lewicki:2021pgr,Laurent:2022jrs}, as well as \cite{Caprini:2019egz, Auclair:2022lcg} for a general discussion). There is no precise prediction for
$v_{\rm w}$ in the 2HDM (or related extensions of the SM) available in the literature.\footnote{See \citere{Dorsch:2016nrg} for estimates
of $v_w$ in the 2HDM for some special parameter configurations.
A simple analytical formula to predict $v_w$ has been found in \citere{Lewicki:2021pgr}. However, this formula has not yet been applied to models with a second Higgs doublet and it is unclear how accurate the prediction for $v_w$ would be for the 2HDM.}
Hence, we will treat $v_{\rm w}$ as a free
parameter in our analysis (see also the discussion in section~\ref{sec:gws}).

\subsection{Vacuum trapping}
\label{sec:vactrapp}
If the condition \eqref{eq:_TN} is not met for any temperature below $T_c$, the Universe would remain stuck in a false vacuum in spite of the existence of a 
\GW{deeper} EW symmetry-breaking minimum of the potential at zero temperature. This phenomenon is dubbed ``vacuum trapping''.
In particular, when aiming to identify the parameter-space regions of a BSM model  where a FOEWPT occurs, the possibility of vacuum trapping highlights that an approach based solely on the critical temperature $T_{c}$ is not sufficient and may yield misleading results. 
Vacuum trapping has been recently discussed in the context of the N2HDM~\cite{Biekotter:2021ysx} and the NMSSM~\cite{Baum:2020vfl} and also previously in the context of color-breaking minima
within the MSSM~\cite{Cline:1999wi}. In the 2HDM, vacuum trapping has been very recently explored in \citere{Goncalves:2021egx}, emphasizing that this phenomenon may take place in particular if the barrier between the false and the true vacua is driven almost exclusively by the radiative corrections, rather than by the thermal contributions to the effective potential.

\subsection{EW symmetry non-restoration}
\label{sec:SnR}

It is known that in certain extensions of the SM the EW symmetry can be broken already at temperatures much larger than the EW scale, resulting in EW symmetry non-restoration
(SnR)~\cite{Espinosa:2004pn, Meade:2018saz, Baldes:2018nel, Glioti:2018roy, Matsedonskyi:2020mlz, Ramsey-Musolf:2017tgh, Carena:2021onl, Biekotter:2021ysx}.
The effect of SnR can exist up to possibly very high temperatures, and it is also possible to find no restoration at all within the energy range in which the model under consideration is theoretically well-defined.
As it has been discussed in \citere{Biekotter:2021ysx} for the 2HDM extended with a real singlet field (N2HDM), in extensions of the SM by a second Higgs doublet
the presence of EW SnR is related to the existence of sizeable quartic scalar
couplings and the impact of the resummation
of infrared divergent modes in the scalar potential.
In that scenario, the maximum temperature up to which the analysis of SnR is valid corresponds to the upper cut-off $\Lambda_{4 \pi}$ defined as the energy scale $\mu$ at which 
one of the quartic couplings reaches the naive perturbative bound $4\pi$. $\Lambda_{4 \pi}$ is
representative of the energy scale $\mu$ at which the theory enters a non-perturbative regime. In section~\ref{secnumanal} we will demonstrate that EW SnR
is possible within the 2HDM, and that SnR appears in regions of the
parameter space adjacent to those where a FOEWPT is present. We will also discuss the
consequences of EW SnR with regards to the viability of a parameter point taking into account the thermal history of the Higgs vacuum, and we briefly comment on the phenomenological implications.\footnote{We make use here of the  numerical treatment of the finite-temperature one-loop effective potential. For a
detailed analytical discussion of SnR in
the high-temperature limit
we refer the reader to \citere{Biekotter:2021ysx}.}

\subsection{Gravitational waves}
\label{sec:gws}
Cosmological first-order phase transitions provide a particularly compelling possibility
for the generation of GWs in the early Universe. The collisions of the expanding bubbles and the resulting motion of the ambient cosmic fluid source a stochastic GW background that could be observable at future GW interferometers (see e.g.~\citeres{Caprini:2019egz,Caldwell:2022qsj,Auclair:2022lcg} for a discussion).~
For FOEWPTs in the 2HDM, where the expanding bubbles do not run-away~\cite{Dorsch:2016nrg}, the contribution from the bubble collisions themselves can be safely neglected. Then, GWs are generated from the 
sound waves and turbulence generated in the plasma
following the bubble collisions~\cite{Caprini:2019egz}.
As introduced in section~\ref{sec:FOEWPT} the GW spectrum produced in a FOEWPT is characterized by four essential quantities~\cite{Caprini:2019egz,Auclair:2022lcg}: the transition temperature $T_*$, the strength $\alpha$, 
the inverse duration of the phase transition $\beta/H$, and the bubble
wall velocity $v_{\rm w}$, i.e.~the speed of the bubble wall
after nucleation in the rest frame of the plasma far away from the phase-transition interface. 
The GW power spectrum as a function of frequency $h^{2}\Omega_{\text{GW}}(f)$ 
is given by
\begin{equation}
h^{2}\Omega_{\mathrm{GW}}(f) =
h^{2}\Omega_{\mathrm{sw}}(f)+
h^{2}\Omega_{\mathrm{turb}}(f) \ ,
\label{eq:GWspectrum}
\end{equation}
where $h^{2}\Omega_{\text{sw}}$ and $h^{2}\Omega_{\text{turb}}$ are the contributions from sound waves and turbulence, \GW{respectively}. 
The contribution from sound waves propagating in the plasma was originally
obtained with the help of large-scale lattice simulations of bubble collisions inducing bulk fluid motion~\cite{Hindmarsh:2017gnf,Hindmarsh:2015qta}.
It can be written as~\cite{Caprini:2019egz} (see also~\cite{Hindmarsh:2016lnk,Hindmarsh:2019phv})
\begin{equation}
\Omega_{\text{sw}}
\left(f\right)=0.687\,
F_{\text{gw},0}\, \Gamma^2 \,\bar U_f^4
\left(\frac{H_{*}R_{*}}{c_{s}}\right)
\tilde{\Omega}_{\text{gw}}
\left(
\frac{H_{*} \tau_{\rm sw}}{c_s}
\right)
C\left(f / f_{\text{sw,p}}\right) \ ,
\label{eq:soundwavesGW}
\end{equation}
with
\begin{equation}
F_{\mathrm{gw},0} = 3.57\cdot 10^{-5}
\left(
\frac{100}{g_*}
\right)^{1/3} \ , \quad
\tilde \Omega_{\rm gw} = 0.012 \ .
\end{equation}
We have also introduced the speed of sound of a relativistic plasma
$c_s = 1 / \sqrt{3}$ and the adiabatic index $\Gamma = 4/3$. $\bar U_f$ is the
the root-mean-square four-velocity of the plasma
given by
\begin{equation}
\bar U_f^2 =
\frac{\kappa \alpha}
{\Gamma (1 + \alpha)} \ ,
\end{equation}
where $\kappa$ denotes the efficiency factor (i.e. the relevant energy
fraction for sound waves), which is a function of $\alpha$ and $v_{\rm w}$ that also depends on the steady-state bubble expansion
regime (deflagrations, detonations or hybrids, see e.g.~\cite{Espinosa:2010hh}), which we obtain following~\citere{Espinosa:2010hh}. The mean bubble
separation $R_*$ entering~\eqref{eq:soundwavesGW} is defined by 
\begin{equation}
H_* R_* =
(8 \pi)^{1/3}
\left(
\frac{\beta}{H}
\right)^{-1}
\mathrm{max}(v_{\rm w},c_s) \ .
\end{equation}
The factor $H_* \tau_{\rm sw}$ in \refeq{eq:soundwavesGW} is introduced in order to account for a timescale $\tau_{\rm sh}$ for the formation of shocks in the plasma (leading to the damping of the sound waves) that may be shorter than one Hubble time~\cite{Ellis:2020awk}
\begin{equation}
H_* \tau_{\rm sw} =
\mathrm{min}(1, H_* \tau_{\rm sh}) \quad
\text{with} \quad
H_* \tau_{\rm sh} \simeq
\frac{H_* R_*}{\bar U_f}
\ .
\end{equation}
Finally, the spectral shape of the sound-wave signal is approximated by the function
\begin{equation}
C(s) = s^3
\left(
\frac{7}{4+3s^2}
\right)^{7/2}
\ ,
\end{equation}
and the associated peak frequency is given by
\begin{equation}
f_{\text{sw,p}} =
26 \left( \frac{1}{H_* R_*} \right)
\left( \frac{T_*}{100\gev} \right)
\left( \frac{g_*}{100} \right)^{1/6}
\mu\mathrm{Hz}
\ .
\label{eq:fpeaksw}
\end{equation}

As indicated above, if the sound-wave period is much shorter
than a Hubble time ($H_* \tau_{\rm sw} \ll 1$), a large fraction of the energy stored in the bulk motion of the plasma does not get to produce GW from sound waves. Yet, when the fluid flow becomes nonlinear (giving rise to shock formation), it can lead to the appearance of turbulence in the plasma, which in turn can also generate GWs. 
Following \citere{Ellis:2020awk}, we have modelled $h^2\Omega_{\rm turb}$ under the most optimistic assumption that all the energy remaining in the plasma when the sound waves are damped gets transformed into turbulence. In this case, the spectrum of GWs from turbulence may be written as~\cite{Caprini:2009yp}
\begin{equation}
\Omega_{\text{turb}} =
7.23 \cdot 10^{-4}
\left(
\frac{100}{g_*}
\right)^{1/3}
v_{\rm w}
\left(
\frac{\beta}{H}
\right)^{-1}
(1 - H_* \tau_{\rm sw} )\,
\Gamma^{3/2}\,
\bar U_f^3\,
D(f,f_{\mathrm{turb},p})
\ ,
\label{omegaturb}
\end{equation}
with peak frequency
\begin{equation}
\frac{f_{\text{turb},p}}{H_*} =
1.75
\left(
\frac{\beta}{H}
\right)
\left(
\frac{1}{\mathrm{max}(v_{\rm w},c_s)}
\right) \mu\mathrm{Hz} \ ,
\quad
\text{with} \quad
H_* =
1.65\cdot 10^{-5}
\left(
\frac{T_*}{100\gev}
\right)
\left(
\frac{g_*}{100}
\right)^{1/6} \ ,
\label{eq:fpeakturb}
\end{equation}
and the spectral shape approximated by
\begin{equation}
D(f,f_p) =
\left(
\frac{f}{f_p}
\right)^3
\left(
1 + \frac{f}{f_p}
\right)^{-11/3}
\left(
1 + 8 \pi \frac{f}{H_*}
\right)^{-1} \ .
\end{equation}
We note in any case that the details of the GW spectrum produced from turbulence constitute a subject of ongoing debate (see e.g.~\citeres{Brandenburg:2017neh,RoperPol:2019wvy,Auclair:2022jod}). At the same time, we have assumed for simplicity in this work that all the energy remaining in the plasma after the sound waves are switched off leads to  turbulence. This gives rise to the factor $(1 - H_* \tau_{\rm sw})$ in \refeq{omegaturb}, to be compared with the factor $H_* \tau_{\rm sw}$ in \refeq{eq:soundwavesGW}. We also stress that the efficiency of turbulence generation as a result of nonlinearities in the plasma is currently under investigation~\cite{Cutting:2019zws}.
Nevertheless, we here find that $\Omega_{\text{turb}}$ plays only
a minor role in our estimate of the GW spectrum, since it has a substantially smaller peak
amplitude compared to $\Omega_{\text{sw}}$.\footnote{In particular, we find that including  $\Omega_{\rm turb}$ does not affect the SNR at LISA at the level of turning an undetectable GW signal into a detectable one. Still, for strong GW signals $\Omega_{\rm turb}$ 
affects the overall GW spectral shape: as will be discussed in more detail in section \ref{sec:2hdm_vactrap_gw}, $\Omega_{\rm turb}$ enhances the signal at the high-frequency tail, which leads to a slight increase in SNR (compared to the GW signal originated
by $\Omega_{\rm sw}$ alone) when the peak frequency of the sound wave contribution $\Omega_{\rm sw}$ is lower than the frequency range for which LISA has the best sensitivity.} 

\vspace{2mm}

The value of the EW scale is such that the GW signal from a FOEWPT
would lie within the frequency sensitivity band of the future space-based LISA GW interferometer.
In order to assess the detectability of a GW signal from a FOEWPT by LISA one has to evaluate the Signal-to-Noise Ratio (SNR) of the GWs.
The SNR can be computed according to~\cite{Caprini:2019egz}
\begin{equation}
\mathrm{SNR}=\sqrt{\mathcal{T}\int_{-\infty}^{+\infty}\text{d}f
\left[\frac{h^{2}\Omega_{\text{GW}}(f)}{h^{2}\Omega_{\text{Sens}}(f)}\right]^{2}} \ ,
\end{equation}
where $\mathcal{T}$ is the duration of the mission times its duty cycle.
We have used values for $\mathcal{T}=3\,\text{y}$ and $7\,\text{y}$. 
$\Omega_{\text{Sens}}(f)$ is the nominal sensitivity of a given LISA configuration to stochastic sources\footnote{
\GW{When} showing the LISA sensitivity curve in this work (e.g.\ in~\reffi{fig9_bis}), it corresponds to the nominal LISA sensitivity $h^{2}\Omega_{\text{Sens}}(f)$ rather than to the so-called power-law sensitivity of LISA~\cite{Thrane:2013oya} to cosmological sources.}, obtained from the power spectral density $S_{h}(f)$
\begin{equation}
h^{2}\Omega_{\text{Sens}}(f)=\frac{2\pi^{2}}{3H_{0}^{2}}f^{3}S_{h}(f),
\end{equation}
with $S_{h}(f)$ taken from the LISA mission requirements~\cite{LISAmissionreq}.
In order to be considered detectable, a GW signal should give rise to
roughly $\mathrm{SNR} \gtrsim 10$~\cite{Caprini:2019egz}.
It should be noted, however, that our model predictions for SNR suffer from
sizable theoretical uncertainties. In particular, both the peak frequency
and the maximum amplitude of the power spectrum $\Omega_{\rm sw}$ depend on the
bubble wall velocity $v_{\rm w}$, for which
no well-established model prediction is available even though there are
promising recent proposals such
as in \citeres{Lewicki:2021pgr,Laurent:2022jrs}.
For most parts of our analysis, we will choose $v_{\rm w} \simeq 0.6$, for which the best prospects regarding GW detection at LISA are obtained in the 2HDM (see section \ref{sec:LISA_colliders} for details).\footnote{Remarkably, in \citere{Laurent:2022jrs} it has been found that for the values of $\alpha$ generically realized in the 2HDM, 
deflagration bubbles with $v_w \sim c_s$ (thus fairly close to our choice $v_w = 0.6$) are 
a relatively common feature of FOEWPTs, independently of the precise microscopic properties of
the BSM model under consideration.}
We nevertheless note that values of $v_{\rm w}$ largely different from $0.6$ may give rise to substantially smaller SNR values at LISA.
Thus, the predictions for the SNRs in our numerical discussion should
be regarded as rough estimates.

\section{2HDM thermal history and phenomenological implications}
\label{secnumanal}

In this section we study the thermal history of the 2HDM regarding a FOEWPT and the associated production of GWs, as well as the occurrence of vacuum trapping
and/or EW SnR. We analyze how these can yield 
\GW{interesting}
constraints on the parameter space of the 2HDM, and we discuss the potential complementarity between collider searches and GW probes with LISA.

\vspace{1mm}

The \GW{possibility of a} FOEWPT in the CP-conserving 2HDM has been extensively studied (see \citeres{Dorsch:2016nrg,Aoki:2021oez,Goncalves:2021egx} for analyses that include a calculation
of the nucleation temperature). The usual scenario that features such a 
first-order transition requires relatively large quartic couplings, which subsequently implies
sizeable splittings among the scalar masses and/or between these masses and the overall (squared) mass scale of the second doublet, $M^{2}=m_{12}^{2}/ s_\beta c_\beta$ \cite{Dorsch:2014qja,Dorsch:2016nrg}. In this work we focus on the 2HDM with type-II Yukawas, for which stringent limits arising from flavor observables constrain the mass of the charged states to be $m_{H^{\pm}} \gtrsim 600 \gev$ \cite{Haller:2018nnx}. This requirement in conjunction with the constraints from electroweak precision observables favors the degeneracy of the masses of the heavy pseudoscalar and the charged scalar, $m_{A} \sim m_{H^{\pm}}$. 
In order to explore the parameter space of the 2HDM taking into account these considerations, we have scanned the parameter space of the
CP-conserving type~II 2HDM over the following ranges of the input parameters,
\begin{align}
t_\beta=3 \ , \quad
m_{h_1} = 125.09 \gev \ , \quad
200 \gev \leq m_{H} \leq 1 \tev \ , \notag \\
600 \gev \leq m_{A} =
m_{H^{\pm}} \leq 1.2 \tev \ , \quad
\cos(\beta-\alpha)=0 , \quad
M^2 = \frac{m_{12}^{2}}{s_\beta c_\beta} = m_{H}^{2} \ .
\label{eqranges1} 
\end{align}
Using \texttt{ScannerS}, we have generated 10k 2HDM
parameter points within the above ranges, passing all the  theoretical and
experimental constraints discussed in section~\ref{sec:treelevel2hdm}.
In a second step, we have analyzed the thermal history of each of these 10k benchmark points with \texttt{cosmoTransitions} \cite{Wainwright:2011kj}, exploring the temperature range $[0, 700 \gev]$. We have studied the temperature dependence of the minima of the one-loop effective potential $V_{\mathrm{eff}}$ from Eq.~\eqref{Veff_Finite} in terms of the two CP-even neutral fields $(\rho_{1}(T), \rho_{2}(T))$. We then have computed the tunneling rate defined in \refeq{eqfuncdeter} between coexisting minima at finite temperature, evaluating whether the criterion from \refeq{eq:_TN} is met and a FOEWPT takes
place.\footnote{We do not take into account the
possibility of CP-breaking or charge-breaking
minima at finite temperature.}

\vspace{2mm}

In section~\ref{sec:2hdm_thhist}, we explore the different thermal histories that the CP-conserving 2HDM features within our parameter scan, which targets the regions where a FOEWPT is realized, as well as the vicinity of such regions.
As mentioned before, a FOEWPT in the 2HDM strongly favours  
sizeable values of the quartic couplings, and we complement this analysis with a study of the energy scale dependence of the quartic couplings. We stress the rich variety of phenomena that arise within this parameter space region, 
and investigate in particular the effects of vacuum trapping and EW SnR.
The analysis of the 2HDM thermal history will allow us to determine the regions of the parameter space in which the strongest FOEWPT can be realized
in the type~II 2HDM, and to assess how strong such transitions are. In section \ref{sec:2hdm_vactrap_gw} we analyze the GW signals that are produced
during the phase transitions. We will compare the
predicted GW signals to the expected LISA sensitivity
in order to assess whether such signals could
be detectable at LISA. Finally, 
in section \ref{sec:LISA_colliders} we compare the prospects of a GW detection at LISA with the collider phenomenology of the corresponding
2HDM parameter 
regions in order to address the question whether those regions could also be probed 
in a complementary way by (HL-)LHC searches.

\subsection{The cosmological evolution of the vacuum in the 2HDM}
\label{sec:2hdm_thhist}

In this section we will investigate possible realizations of non-standard
cosmological histories in the 2HDM. Even though the motivation for the
analyzed parameter plane was its suitability for the occurrence of FOEWPTs, as described above, we 
point out that the considered parameter space also features a
rich variety of thermal histories in terms of the
patterns of symmetry breaking and symmetry restoration.

\begin{figure}[t]
\centering
\includegraphics[width=0.6\textwidth]{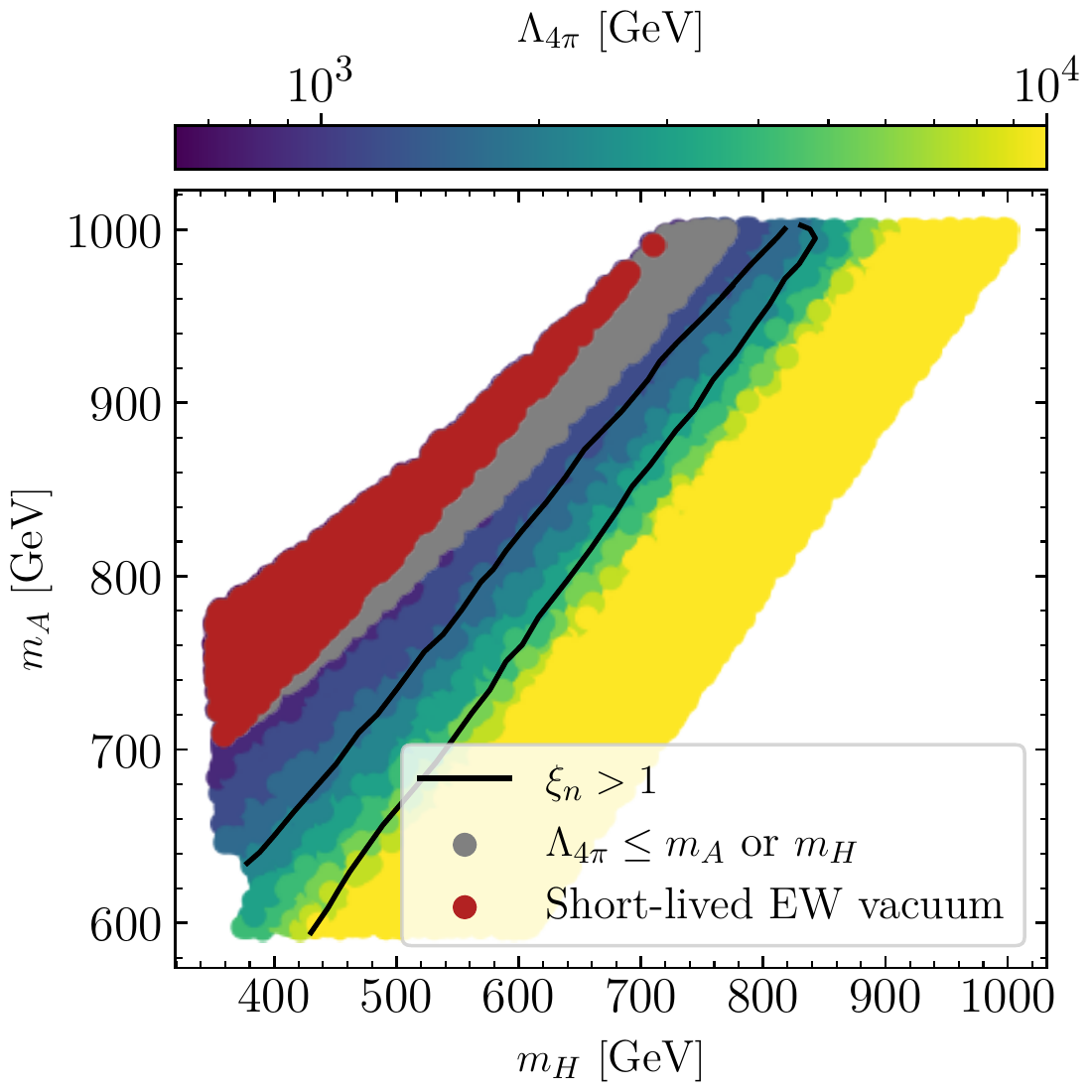}
\caption{\small 
Constraints from perturbativity and vacuum stability,
and region featuring a strong FOEWPT in the plane of the mass of the heavy CP-even scalar $m_{H}$ and the masses of the CP-odd scalar and the charged scalars 
$m_{A} = m_{H^\pm}$ in the type~II 2HDM, with the other parameters specified in \refeq{eqranges1}. The displayed points pass all the theoretical and
experimental constraints discussed in section \ref{sec:treelevel2hdm}.
The color bar indicates the energy scale $\Lambda_{4 \pi}$ at which one of the quartic couplings of the parameter point reaches the naive perturbative bound $4 \pi$ (for points with $\Lambda_{4 \pi}< 10 \tev$). Points with $\Lambda_{4 \pi} < m_{A}$ or $m_{H}$ are indicated in gray, and points with a short-lived EW
vacuum are shown in red. Yellow points feature $\Lambda_{4 \pi} \geq 10 \tev$.
The black line circumscribes all the points that feature a strong
FOEWPT (see text for details).}
\label{Fig1}
\end{figure}

Before we start the discussion of the 2HDM cosmological history,
we briefly inspect the additional constraints from the RGE running of
the parameters, that we have applied in order to restrict the analysis to
parameter benchmarks for which our perturbative analysis is applicable.
Since we are interested in FOEWPTs, we explore a parameter space region where
relatively large quartic couplings are present.
A key check on the validity of our perturbative calculation 
of the quantities that characterize the FOEWPT
is to make sure that at the energy scales relevant for our analyses the values of the couplings remain in the perturbative range $|\lambda_i| < 4 \pi$ (see section \ref{sec:rges} for details).
In \reffi{Fig1} we show the analyzed parameter space in the $(m_H, m_A)$ plane of the 2HDM of type~II as specified in \refeq{eqranges1}. For each point we indicate the energy scale $\Lambda_{4 \pi}$ at which one of the 2HDM quartic couplings
reaches the naive perturbative bound $4\pi$. 
The lower-right corner in which no points are shown is excluded from
the requirement on the tree-level potential to be bounded
from below, imposed via 
\texttt{ScannerS}.\footnote{Such parameter points could still feature
a bounded potential upon inclusion of loop corrections~\cite{Staub:2017ktc}.
We did not include this possibility
in our analysis because we focus here on the thermal evolution of the potential. Including the boundedness check for the loop-corrected
scalar potential at zero temperature
is computationally much more
expensive compared to
the application of the tree-level conditions
which were determined in compact
analytical form~\cite{Barroso:2013awa}.}
In the lower right strip we find points with $\Lambda_{4 \pi} \geq 10 \tev$, which are indicated in yellow.
On the other hand, we find that a large part of the 
parameter space that is allowed by the constraints discussed in section \ref{sec:treelevel2hdm} features relatively low values for $\Lambda_{4 \pi}$, smaller than $10 \tev$. This feature arises as a consequence of the sizeable
values of the quartic couplings $\lambda_{i}$ at the initial scale $\mu_{0} = v$ that are required to achieve large splittings among the scalar masses, as described in section \ref{secnumanal}. 
In particular, our scan contains points for which $\Lambda_{4 \pi} < m_{A}=m_{H^{\pm}}$ or $\Lambda_{4 \pi} < m_{H}$, which are shown in gray in \reffi{Fig1}. Since for
these points the perturbativity bound is reached for an energy scale that is lower than one of the involved masses, we regard such a situation as 
unphysical. Accordingly, we consider this parameter region as 
excluded and will not analyze it further.
As will be discussed below, this region exclusively features
scenarios where the global minimum of the potential at $T=0$ is the origin of field space.  Consequently, this
additional constraint does not exclude parameter points that otherwise would predict
a FOEWPT. Furthermore, we verified that a subset of points with $\Lambda_{4 \pi} < m_{A}=m_{H^{\pm}}$ or
$\Lambda_{4 \pi} < m_{H}$ features a short-lived EW minimum, 
i.e.\  the probability for quantum tunnelling from the 
EW minimum into the deeper minimum (the origin of field space) in this case is such that it gives rise to a lifetime of the EW vacuum that is substantially smaller than the age of the universe.\footnote{The calculation of the lifetime of the
EW vacuum relies on the computation of the four-dimensional euclidean
bounce action instead of the three-dimensional bounce action that determines
the decay rate of the false vacuum at finite temperature.
It should also be noted that in the scenario investigated
here the presence of the global minimum in the origin only arises at the loop level,
such that a tree-level analysis of the EW vacuum stability would not be sufficient
here.}
The points with a short-lived EW vacuum
are shown in red in \reffi{Fig1}. Finally, all the points that feature a strong
FOEWPT in \reffi{Fig1} are circumscribed by a solid black line. The strong FOEWPT region is characterised by
\begin{equation}
     \xi_{n}=\frac{v_{n}}{T_{n}}>1,
\label{sFOEWPT}
\end{equation}
where $v_{n}$ is the vev in the minimum adopted by the Universe at the nucleation temperature $T_n$. We stress that for values of $\xi_{n}$ substantially smaller than 1 it becomes numerically impossible to distinguish between a first- and a second-order phase transition in a perturbative analysis, and such a distinction would then require
to take into account non-perturbative effects~\cite{Kajantie:1996mn,Niemi:2020hto}.

\vspace{2mm}

\begin{figure}
\centering
\includegraphics[width=0.6\textwidth]{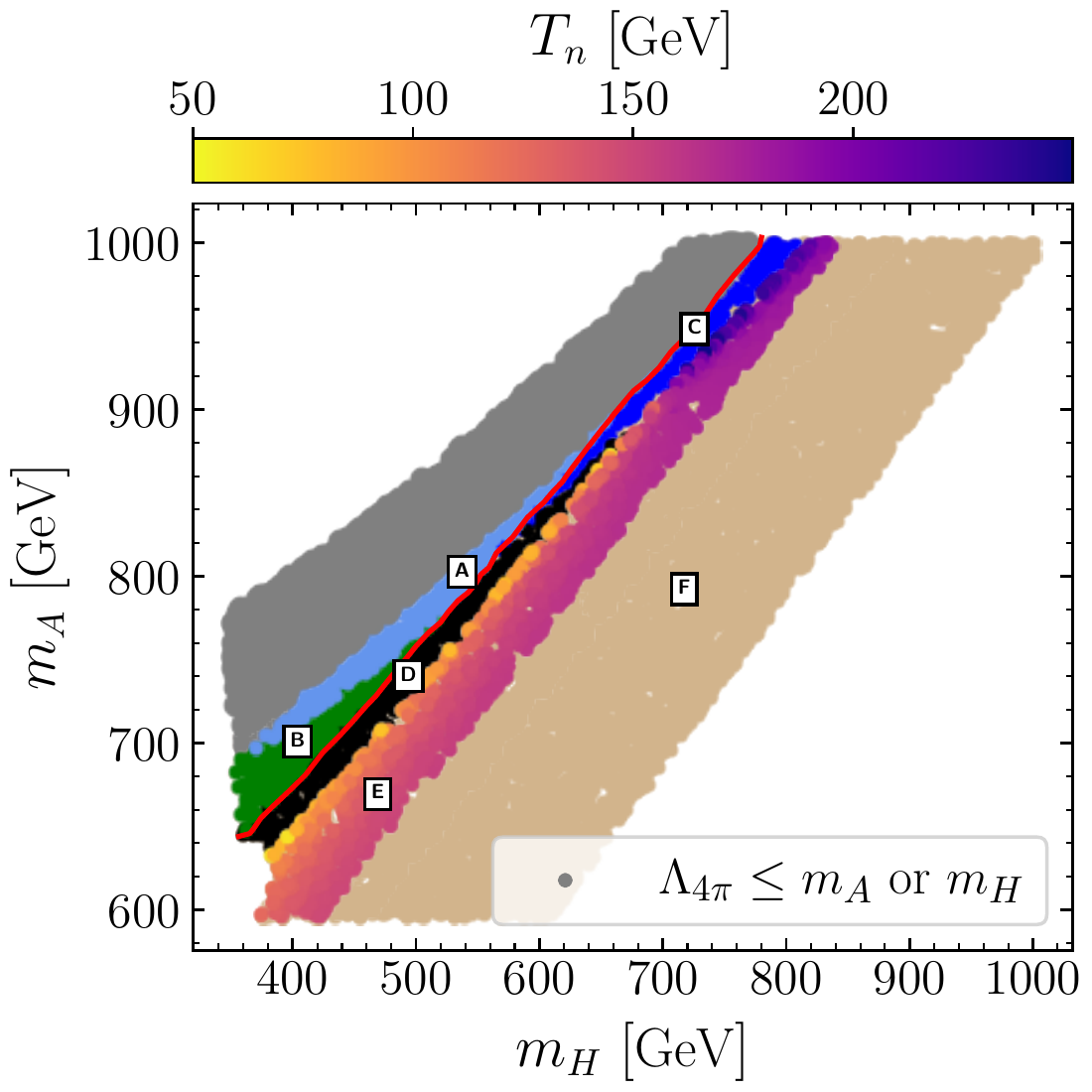}
\includegraphics[width=0.85\textwidth]{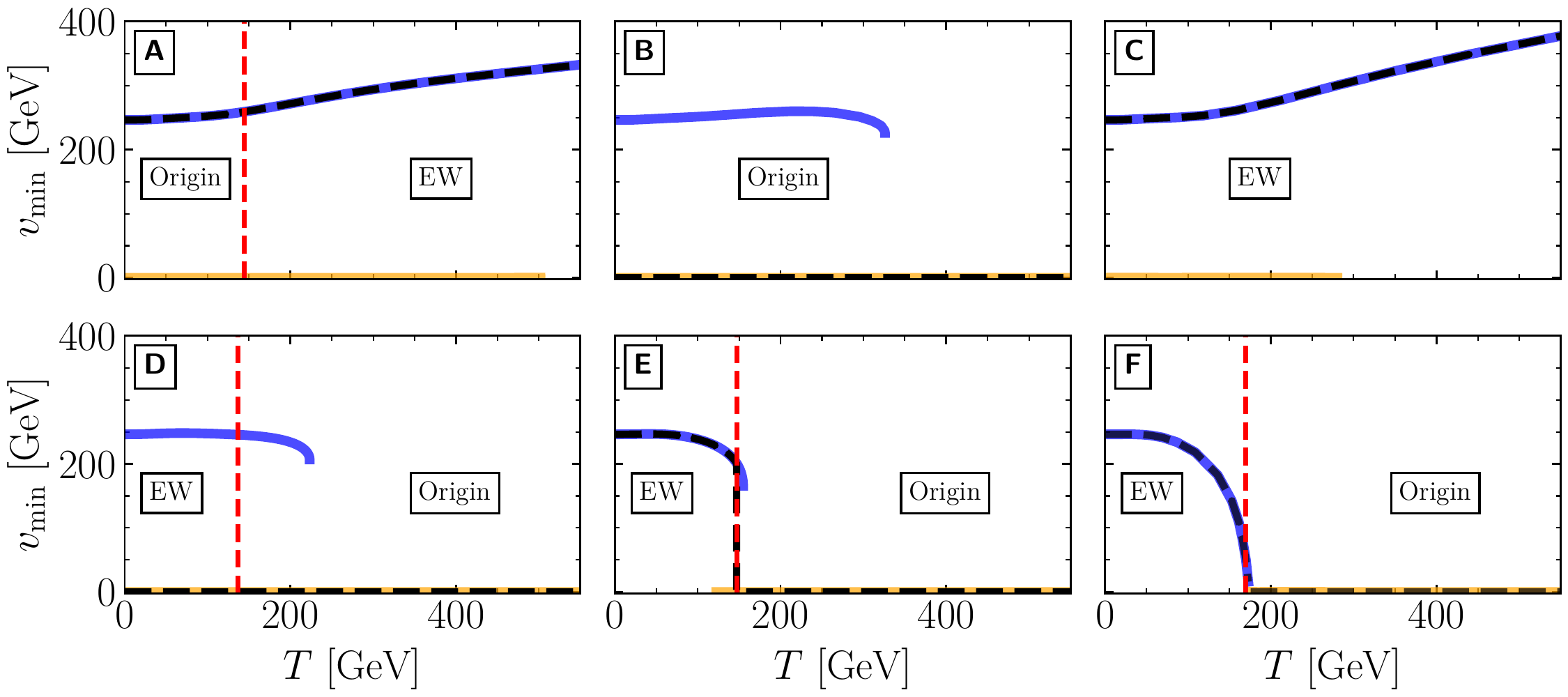}
\caption{Top: 
The parameter plane as shown in \reffi{Fig1}, with zones featuring qualitatively different thermal histories of their vacuum structure labelled as A, B, C, D, E and F. The red line separates the region with a zero-temperature global minimum at the origin of field space (left) from the region with a zero-temperature electroweak global minimum (right). Bottom: characteristic temperature dependence of $v_{\rm min}$ for the local minima of the potential for each
of the six labelled regions. The blue lines indicate the temperature evolution of 
$v_{\text{min}}$ evaluated at the minimum where the electroweak symmetry is 
broken. The orange lines denote how the minimum where the electroweak symmetry is 
unbroken evolves. The dashed black lines show the vacuum configuration
adopted by the universe taking into account phase transitions between co-existing
minima. The vertical red lines show the critical temperature, and the labels ``origin'' and ``EW'' indicate the global minimum of the potential. 
}
\label{fig2}
\end{figure}
We now discuss the different kinds of symmetry-breaking patterns that
occur in the analyzed parameter space. In the upper plot of \reffi{fig2}, we indicate six qualitatively distinct zones of the $(m_H, m_A)$ plane of the 2HDM of type~II shown in \reffi{Fig1}, labelled by A, B, C, D, E and F (as discussed above, in our analysis we regard the gray/red points as excluded).
Each of the six zones features a different temperature evolution of the vacuum configuration of the 2HDM Higgs potential.
The red line divides the mass plane into two regions. The points above and to the left
of the red line feature at $T = 0$ a global minimum at the origin of field space,
whereas those 
below and to the right of the red line have the EW minimum as global minimum at $T = 0$. The different zones in the upper plot of \reffi{fig2} are analyzed individually in the six plots shown in the lower part. These plots indicate the typical temperature dependence of the minima of the potential for each 
of the six labelled regions (where the specific point is taken where the label is located). The six benchmark points have been analyzed with \texttt{cosmoTransitions} up to a temperature $T_{\rm max}=550 \gev$. The blue lines indicate the temperature evolution of $v_{\text{min}} \equiv \sqrt{v_{1}^2 + v_{2}^2}|_{\text{min}}$ 
evaluated at the minimum where the electroweak symmetry is broken. The absence of a blue line for a given temperature indicates that no EW symmetry breaking minimum exists at this temperature. The orange line shows the temperature dependence of the minimum located at the
origin of field space. The absence of this line for a given temperature shows that there is no (local or global) minimum at the origin of field space.
The vertical dashed red lines show the temperature at which the two minima involved in the transition are degenerate, i.e.~the critical temperature. The label ``origin''
corresponds to a range of temperatures where the origin is the global minimum of the potential, and ``EW'' indicates a global minimum where the EW symmetry is broken.
Taking into account the possible transitions between coexisting minima, the 
dashed black line indicates the temperature dependence of the vev actually adopted by the universe for each of the benchmark scenarios.

The parameter points with a zero-temperature global minimum at the origin, i.e.\ the points on the upper left of the red line, are classified into two different zones (A and B). We find that a zero-temperature vacuum stability analysis would allow those points as they all feature meta-stable EW minima whose lifetime is
compatible with the age of the universe.
The benchmark point belonging to zone~A
has an EW-broken minimum for the entire temperature range explored, whereas a minimum at the origin only appears for temperatures below $T \sim 500 \gev$. Consequently, the adopted vacuum configuration at high temperature is the one breaking the EW symmetry, and zone A features
EW SnR at high temperature. 
This implies that the breaking of the EW symmetry in the early Universe would have taken place at temperatures
substantially above the EW scale (in particular $T > T_{\rm max}$).
Such a high value of the transition temperature can have profound consequences in the context of EW baryogenesis and the related phenomenology at colliders or other low-energy
experiments searching for CP-violating effects.
In view of those features and of the existing limits on BSM physics around the
EW scale at the LHC, the proposal of EW \textit{high-scale} baryogenesis
has gained attention in recent years~\cite{Baldes:2018nel,Meade:2018saz,
Glioti:2018roy,Matsedonskyi:2020mlz,Matsedonskyi:2020kuy,Bodeker:2020ghk,
Biekotter:2021ysx,Carena:2021onl,Matsedonskyi:2021hti}.
Based on the perturbative treatment of the effective potential,
we find in this work that the 2HDM, or more broadly speaking extensions of the SM containing
a second Higgs doublet, could feature EW SnR and possibly allow for EW baryogenesis at energy scales much higher than the EW scale.
On the other hand, for the benchmark scenario belonging to zone~B, the only existing minimum at $T_{\text{max}}$ is the minimum at the origin, i.e.\ the EW
symmetry is restored at the maximum temperature that we have analyzed.
The broken phase appears for temperatures below $T \sim  325 \gev$, but never becomes deeper than the minimum
at the origin, which remains the global minimum for all~$T$.
A phase transition into the broken phase is not possible, and the EW
symmetry is preserved as the temperature
approaches zero. Consequently, this parameter region is
regarded as unphysical and therefore excluded.

\vspace{2mm}

Now we turn to the analysis of the parameter space region that features a global EW minimum at $T = 0$, located on the lower right side of the red line in the upper plot of \reffi{fig2}. Here we identify four qualitatively different
zones depending on their thermal histories (C, D, E, F). 
For the benchmark point of region C, 
an EW symmetry breaking minimum exists
already at $T_{\rm max}$, whereas no minimum of the potential at the origin exists at this temperature. Consequently, this zone exhibits EW SnR
at high temperature. The EW minimum is always deeper than the one at the origin, which for our chosen benchmark within this region appears for temperatures below $T \sim 280 \gev$, such that no transition to the minimum at the origin
can occur, and the parameter points in this
region are, at least in principle, not excluded (in order to definitely determine whether such points are physically viable, one would require a detailed analysis of the behaviour of the scalar potential at even higher temperatures).

Region~D features the phenomenon of vacuum trapping. In the benchmark scenario shown in plot~D, the EW symmetry is restored at high temperature,
and the EW phase appears for temperatures below $T \sim 225 \gev$.
Even though a critical temperature exists in this scenario, 
the condition \refeq{cond_trans} is never satisfied,
and as a consequence the universe remains trapped in a false vacuum at the origin as $T \rightarrow 0$. This parameter region is therefore not phenomenologically viable and has to be excluded.
The possibility of vacuum trapping in the thermal history of the universe and its phenomenological implications will be further discussed in section \ref{sec:2hdm_vactrap_gw}.

All the points in region~E feature a strong FOEWPT, where the 
quantity $\xi_{n}$ meets the condition \eqref{sFOEWPT}. The plot~E exemplifies the typical temperature dependence of the vacuum configuration 
for one of such parameter points. In this benchmark scenario, the EW
symmetry is restored at $T_{\rm{max}}$. The EW minimum appears for temperatures below $T \sim 155 \gev$, and  a strong FOEWPT takes place at a nucleation temperature $T_{n} \approx 140 \gev$. The nucleation temperatures for all points in zone~E are given by the color coding in the upper plot of \reffi{fig2}. 
In region~E gravitational wave signals that are
sufficiently strong to be detected by LISA could potentially be 
generated. In section \ref{sec:gravitationalwaves}, we 
will discuss zone~E regarding the possible detectability of such GW signals by LISA.

Finally, the points in zone~F feature either a weak FOEWPT with $\xi_n < 1$ or a second-order EW phase transition.\footnote{The numerical precision of the calculation of $\xi_n$ is not sufficient to distinguish between a very weak FOEWPT, $\xi_n \ll 1$, and a second-order EW phase transition, but for the purpose of our paper
such a distinction is of no phenomenological relevance anyway.} The 
plot~F shows a specific benchmark in this region with a second-order phase transition (or a very weak FOEWPT) taking place at $T \sim 170 \gev$.
At low temperature the minimum adopted by the universe breaks the EW
symmetry, whereas the minimum adopted at 
high temperature is located at the origin of field space and therefore the EW symmetry is restored.

\vspace{2mm}

To summarize the above discussion, taking into 
account the requirement that the universe has to reach the 
correct minimum that breaks the EW symmetry at zero 
temperature has shown that the regions B and D are 
unphysical and have to be excluded.

\subsection{Phenomenological consequences of vacuum trapping}
\label{sec:2hdm_vactrap_gw}

Vacuum trapping, as outlined in section~\ref{sec:vactrapp}, corresponds to 
the situation where the universe remains trapped in an EW symmetric phase 
while it cools down, even though a 
\GW{deeper} EW symmetry breaking minimum of the
potential exists at zero temperature. The potential in this case is such that
\refeq{cond_trans} is never fulfilled at any temperature at which the EW symmetry breaking minimum is deeper than the minimum at the origin.\footnote{We stress that in the 2HDM analysis presented in this paper we did not encounter vacuum trapping in any false minimum other than the one located at the origin.}
Several recent analyses~\cite{Baum:2020vfl,Biekotter:2021ysx, Goncalves:2021egx}
have noted the importance of this phenomenon for the phenomenology of models with 
extended Higgs sectors, in particular regarding the possibility of a FOEWPT,
the realization of EW baryogenesis, or the production of a stochastic GW background. 
As we will show in the following, taking into account the constraints from vacuum trapping has an important impact on the prospects for probing parameter regions featuring such phenomena at particle colliders. 
We start with an analysis of the implications of vacuum trapping for parameter regions in which EW baryogenesis could occur.
Afterwards we discuss the impact of vacuum trapping on the possibility of generating
GW spectra during a FOEWPT in the 2HDM with a sufficient amplitude to be 
detectable at future GW observatories.

\subsubsection{Implications for electroweak baryogenesis}
\label{sec:bau}

Although the LHC has set important limits on the presence of additional Higgs bosons at the EW scale, the 2HDM remains compatible with those limits as a viable framework
for the explanation of the matter--antimatter asymmetry of the Universe by means of
EW baryogenesis~\cite{Dorsch:2016nrg}.
In addition to new sources of CP-violation that can be present in the 2HDM compared 
to the SM, another vital ingredient for the realization of baryogenesis
is the presence of a strong FOEWPT. In the following, we will focus on 
the criterion of a FOEWPT.\footnote{We assume that the required sources of CP violation do not have an impact on the dynamics of the phase transition and can therefore be neglected in our analysis.} As an indicator of the presence of a FOEWPT that is sufficiently strong for allowing the generation
of the observed matter--antimatter asymmetry, the criterion
\begin{equation}
     \xi_{c}=\frac{v_{c}}{T_{c}}>1,
\label{sFOEWPTTc}
\end{equation}
has often been used in the 2HDM and extensions thereof~\cite{Cline:2013bln,Dorsch:2014qja,Basler:2016obg,Basler:2017uxn,Dorsch:2017nza,Bernon:2017jgv,Basler:2019iuu,Fabian:2020hny,Su:2020pjw,Gabelmann:2021ohf,Atkinson:2021eox,Atkinson:2022pcn}.
Here $v_{c}$ is the vev in the EW symmetry breaking minimum at the critical temperature $T_c$, and $\xi_{c}$ is denoted as the \textit{strength}
of the transition. This so-called baryon number preservation criterion~\cite{Patel:2011th} (see also \citere{Quiros:1999jp} and references therein) yields a condition for avoiding the wash-out of the baryon asymmetry after the EW phase transition. 
However, in parts of the literature it is also used as a sufficient
requirement for the presence of a FOEWPT via the existence of the critical temperature $T_c$ at which the EW minimum becomes the global minimum.
In contrast to this, we will show in this section that the criterion of \refeq{sFOEWPTTc} is not a reliable indicator of the occurrence of a FOEWPT in the 2HDM  (see also \citere{Patel:2011th}).
As analyzed below, 
the calculation of the nucleation (or transition) temperature with the help of \refeq{cond_trans} is crucial, not only in order to assess the actual strength of the FOEWPT which happens at temperatures $T_* \sim T_n < T_c$, but more importantly to determine whether the FOEWPT takes place at all.
The nucleation criterion shown in \refeq{cond_trans} should then be used in order to accurately determine the 2HDM parameter space that reaches the EW vacuum configuration at zero temperature as a result of a FOEWPT, whereas a criterion \GW{just} based on the existence of $T_c$ would include also parameter space regions that are unphysical due to the occurrence of vacuum trapping.

\begin{figure}[t]
\centering
\includegraphics[width=0.48\textwidth]{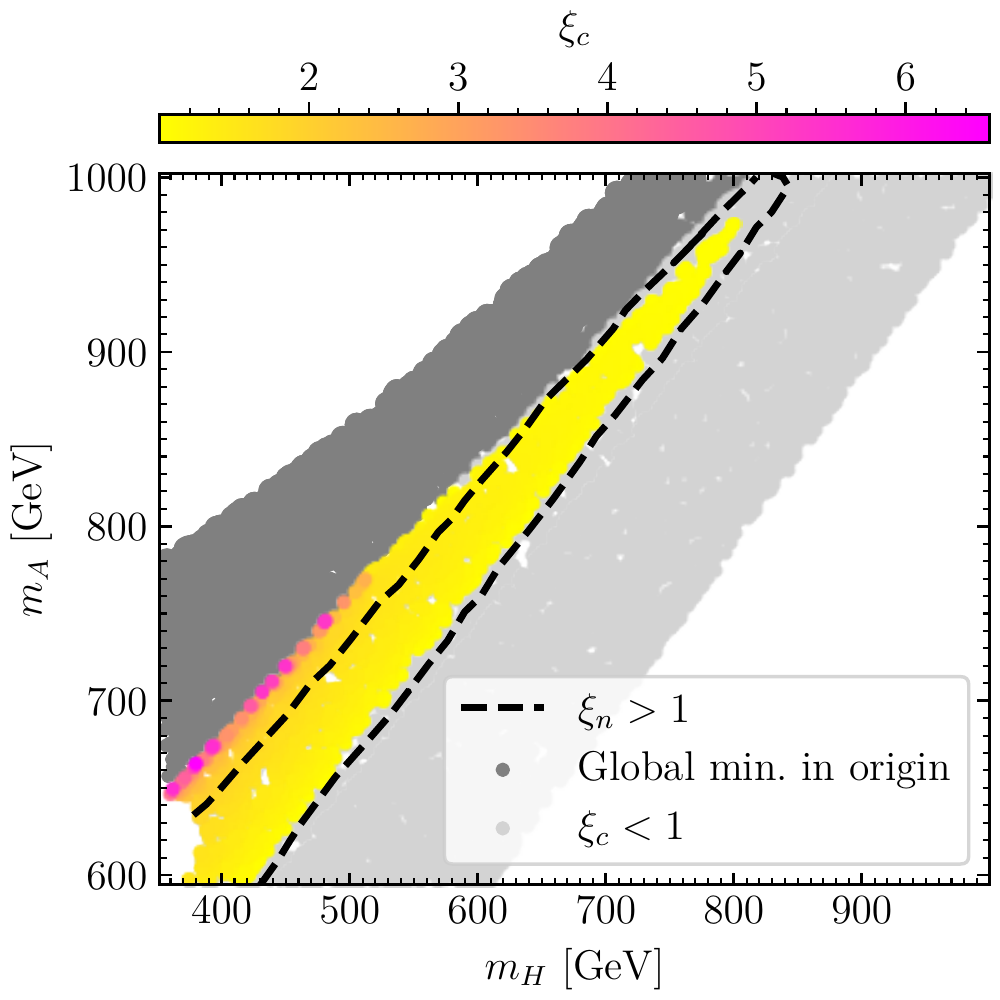}~
\includegraphics[width=0.48\textwidth]{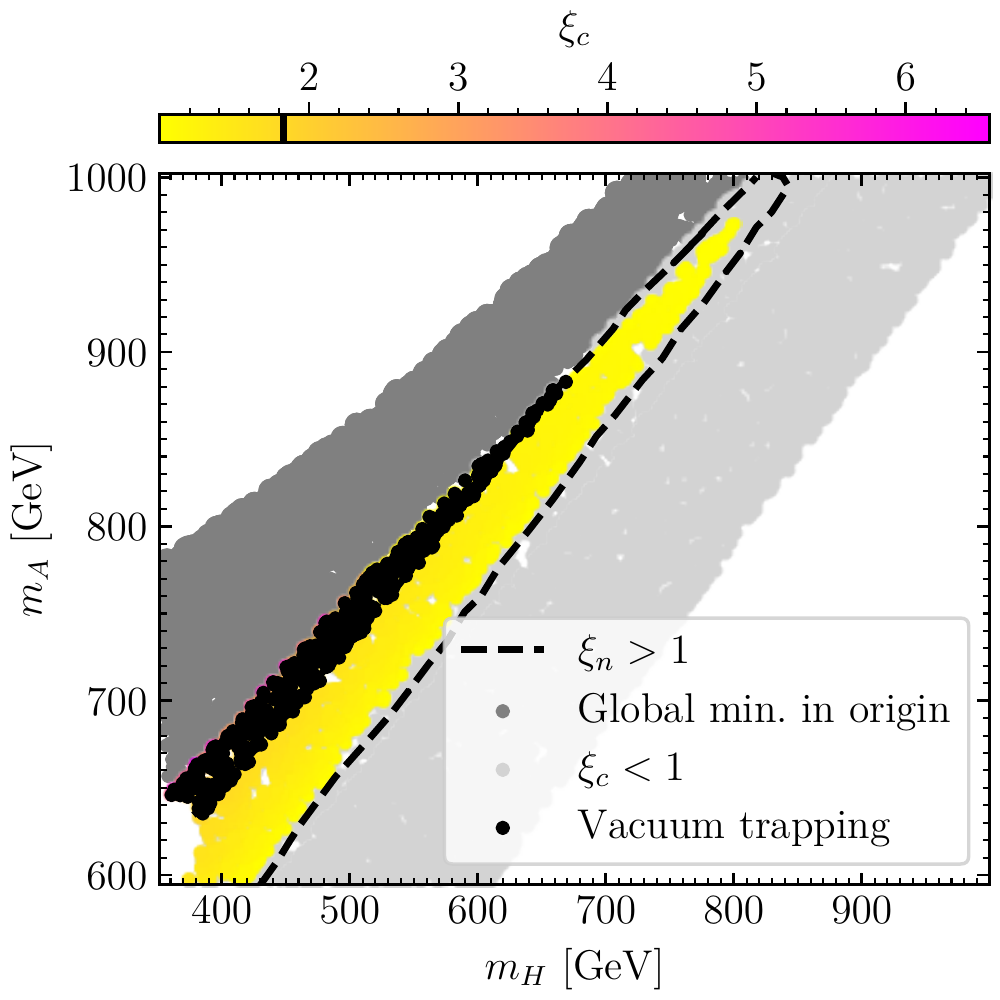}
\caption{\small 
The parameter plane as shown in \reffi{Fig1}, where \GW{for both plots}
the points shown in light gray feature a second-order EW phase transition
or a FOEWPT with $\xi_{c} < 1$, whereas for the
dark gray points the global minimum is in the origin
(corresponding to the area of the gray points
and the zones~A and ~B in \reffi{fig2}), and accordingly the points do not feature
an EW phase transition within the investigated temperature range. The colored points feature a critical temperature $T_c$ at which the EW minimum becomes the global one, where the color coding of the points indicates the value of $\xi_{c}$. The dashed black line circumscribes all points that feature a FOEWPT
with $\xi_{n} > 1$. 
In \GW{addition to what is shown in the left plot, the black points 
in the right plot (which are painted above the points displaying the value of $\xi_{c}$)}
indicate the parameter region that is excluded 
as a consequence of vacuum trapping, and the vertical black line
in the color bar indicates the maximum value of $\xi_{c}$ that is found after the 
incorporation of the constraint from vacuum trapping.}
\label{fig3}
\end{figure}

In \reffi{fig3} we show the parameter scan points in the $(m_H, m_A)$
plane, where the color coding indicates (\GW{for both} plots) the values of $\xi_{c}$ for parameter points for which $\xi_{c} > 1$. According to several existing analyses (see the discussion above) these points would be classified as featuring a strong FOEWPT that could generate the observed baryon asymmetry of the Universe.
The dark gray points in \reffi{fig3} correspond to the region with a zero-temperature
global minimum at the origin of field space (corresponding in \reffi{fig2} to the combined area of the gray points and of the zones~A and~B). 
These points are thus not relevant for the present analysis (being either unphysical or featuring EW SnR up to the highest temperatures analyzed in our scan).
The light gray region depicts parameter points that, while featuring a zero-temperature global EW minimum, do not meet the condition imposed on the strength of the transition based on $T_{c}$, see \refeq{sFOEWPTTc}. The dashed black line circumscribes the points that meet the more appropriate requirement for a strongly FOEWPT based on $T_{n}$, defined in \refeq{sFOEWPT} (coinciding with the solid black line in \reffi{Fig1} and the zone~E in \reffi{fig2}).
\GW{The left plot of} \reffi{fig3} shows that the region with the highest values of $\xi_{c}$ (corresponding to the pink points) lies at the border with the dark gray region, 
and features transition strength values up to $\xi_{c} \sim 6$, which would be particularly well suited for EW baryogenesis.
However, taking into account the constraint from vacuum trapping (zone~D in \reffi{fig2}), indicated by the black points in \GW{the right plot of \reffi{fig3}, which are painted above the points displaying the value of $\xi_{c}$}, one can see that the parameter region featuring the highest $\xi_{c}$ values is in fact excluded as a consequence of vacuum trapping. After taking into account this constraint, the maximum allowed value for $\xi_{c}$ 
is $\xi_{c} \sim 1.8$ (instead of $\xi_{c} \sim 6$), indicated by a vertical black line inside the color bar on the right plot of \reffi{fig3}.
At the same time, \reffi{fig3} highlights that vacuum trapping not only has a strong impact on the maximum values of $\xi_{c}$ that can be achieved in the physically viable parameter regions,
but it is also crucial for determining the 2HDM parameter region that \GW{features} a FOEWPT: the constraint from vacuum trapping excludes the parameter region in \GW{the left plot of} \reffi{fig3} with the largest values for the mass splitting $m_A - m_H$ for a fixed value of $m_H$. 
This has important consequences for the prospects of probing 2HDM scenarios 
featuring a strong FOEWPT at current and future colliders. For instance, the cross section for the LHC signature $pp \to A \to Z H$ (which would 
\GW{be a ``smoking gun'' signature for}
the existence of such a strong FOEWPT in the 2HDM~\cite{Dorsch:2014qja,Dorsch:2016nrg,Dorsch:2017nza})
depends on the mass splitting between $A$ and $H$, since the branching ratio for the decay $A \to Z H$ grows with increasing mass splitting. 
The constraint from vacuum trapping can then place an
upper limit on the cross section for such an $A\to Z H$ signature within the 2HDM (see e.g.~\cite{Goncalves:2021egx}).
A more detailed discussion on the collider phenomenology of the parameter region with a FOEWPT will be given in section \ref{sec:LISA_colliders}.

Finally, we point out that the black-dashed line in \reffi{fig3},
defined by the criterion $\xi_{n} > 1$, circumscribes also light-gray
points at the upper end of the $m_A$, $m_H$ mass ranges considered here. Thus, in this mass region we find parameter points that feature strongly
FOEWPTs based on the transition strength evaluated at $T_n$, but would not satisfy the corresponding criterion for avoiding the wash-out of the baryon asymmetry evaluated at $T_c$. As a consequence, the criterion based on $T_n$ allows for larger values of $m_A$ and $m_H$  compared to the (potentially misleading) criterion based on $T_c$.

\subsubsection{Gravitational waves}
\label{sec:gravitationalwaves}

As discussed in section \ref{sec:gws}, a cosmological FOEWPT can produce
a stochastic GW background that could be observable by the future LISA GW interferometer. We now analyze the production of GWs from a FOEWPT in the 2HDM, discussing the quantities $T_{*}$, $\alpha$, $\beta/H$ and $v_{w}$ and studying the prospects for the detection of the GW signals at LISA. 
We will specifically show 
\GW{that} the phenomenon of vacuum trapping
puts severe limitations on the GW SNR achievable at LISA in the 2HDM.

  \begin{figure}[t]
    \centering
    \includegraphics[width=0.65\textwidth]{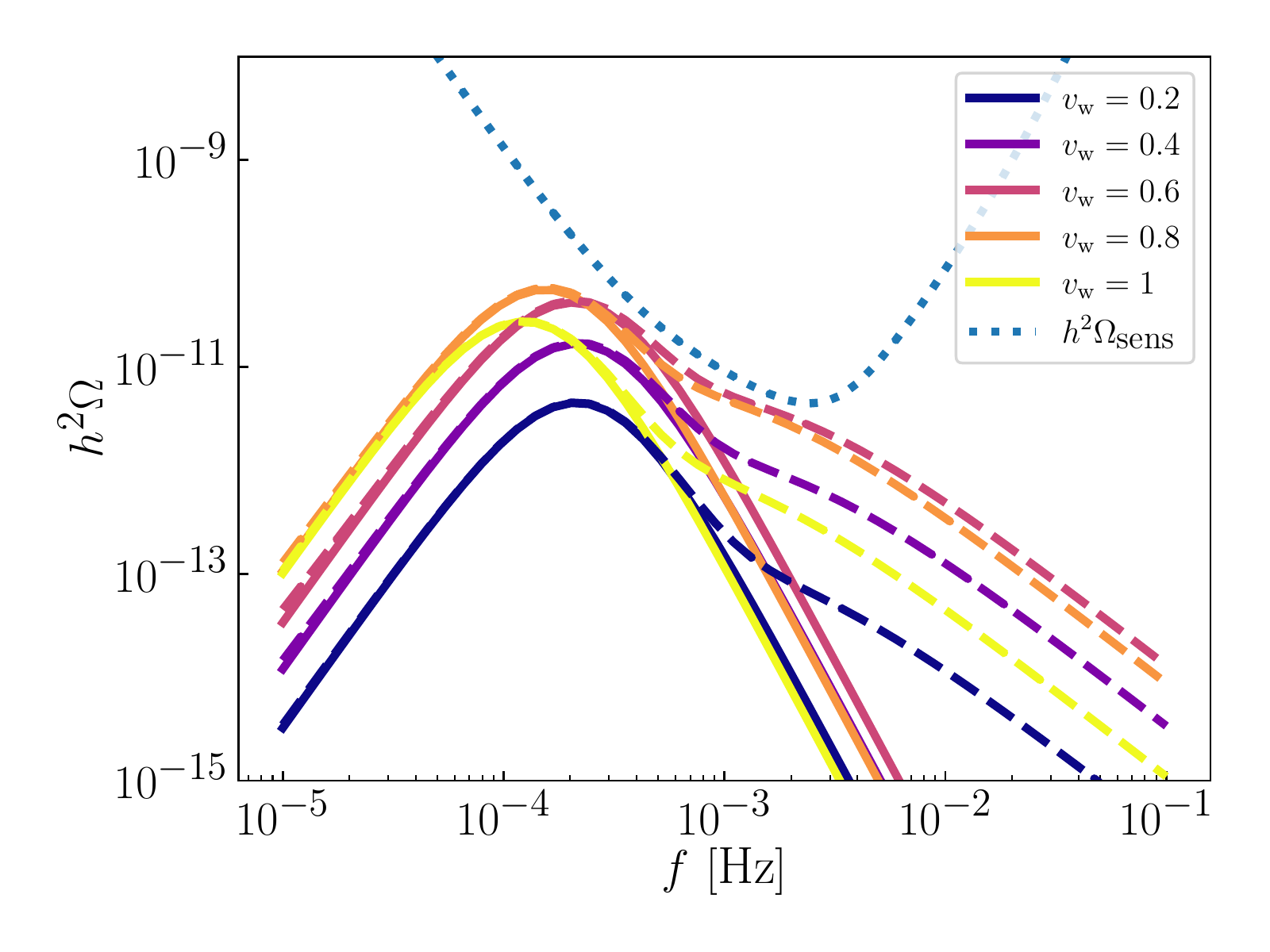}
        
    \vspace{-3mm}
    
    \caption{ \small GW spectrum for a 2HDM benchmark point with BSM scalar masses $m_{H}=419.33 \gev$ and $m_{A}=m_{H^\pm}=663.05 \gev$, yielding a FOEWPT with $T_n =52.43 \gev$,
$\alpha = 0.172$ and $\beta / H = 26.2$. $h^{2} \Omega_{\rm GW}$ predictions for different bubble wall velocity values ($v_{\rm w} = 0.2,0.4,0.6,0.8,1$) are shown in different colors for the concave curves. 
\GW{The dashed lines show the predictions where the turbulence contribution to $h^{2} \Omega_{\rm GW}$ is included, while this contribution is omitted for the solid lines.
The dotted curve indicates the nominal sensitivity of LISA to stochastic sources, $h^{2} \Omega_{\rm sens}$.}} \label{fig9_bis}
\end{figure}

We first briefly discuss the dependence on the bubble wall velocity $v_{\rm w}$. 
In \reffi{fig9_bis} we show, for different values of $v_{\rm w}$,  the predictions for the GW spectrum of a specific 2HDM benchmark point \GW{yielding a potentially large GW signal} with BSM scalar masses $m_{H}=419.33 \gev$ and $m_{A}=m_{H^\pm}=663.05 \gev$, yielding a FOEWPT at a temperature of $T_n =52.43 \gev$ with $\alpha = 0.172$ and $\beta / H = 26.2$.
The solid lines correspond to the predictions for $h^2 \Omega_{\rm GW}$ omitting the contribution from turbulence in the plasma, whereas \GW{for} the dashed lines \GW{this contribution is included}. \reffi{fig9_bis} also shows the LISA nominal sensitivity obtained from its noise curve (see section~\ref{sec:gws} for details). The bubble wall velocity has a strong impact on the GW spectrum, shifting the position of the peak of the GW signal and significantly modifying its amplitude. These \GW{effects} translate into a large variation of the SNR at LISA (assuming a duration of the LISA mission of $\mathcal{T} = 7$ years) for different values of $v_{\rm w}$, as shown in Table~\ref{Table1_SNR}. 
The highest SNR occurs for $v_{\rm w} \sim 0.6$ \htb{and for GW signals in which the turbulent motion of the primordial plasma was considered as a source of GWs}. \GW{The feature that the highest SNR occurs for about this value of $v_{\rm w}$} is fairly generic in the 2HDM (\GW{i.e.\ it is not specific for} the benchmark chosen for illustration). \htb{We thus consider the contribution from turbulence to $h^2 \Omega_{\rm GW}$ and} use $v_{w}=0.6$ for the predictions of the GW signals in the rest of this work \GW{in order to investigate the maximum sensitivity to these signals}.

\begin{table}
\centering{}
    \begin{tabular}[b]{ccc}\hline
      $v_{\rm w}$ & turb. & no turb. \\ \hline
      0.2 & 23 & 18 \\
      0.4 & 149 & 67 \\
      0.6 & 522 & 153 \\
      0.8 & 431 & 101 \\
      1 & 70 & 28 \\
       \hline
    \end{tabular}
            \caption{LISA SNR of the GWs for the 2HDM benchmark scenario shown in~\reffi{fig9_bis} for different values of the bubble wall velocity $v_{\rm w}$, taking into account the effect of turbulence as a source of GWs ("turb.") and 
        \GW{omitting}
    it ("no turb.").}
    \label{Table1_SNR}
  \end{table}

\begin{figure}[t]
\centering
\includegraphics[width=0.55\textwidth]{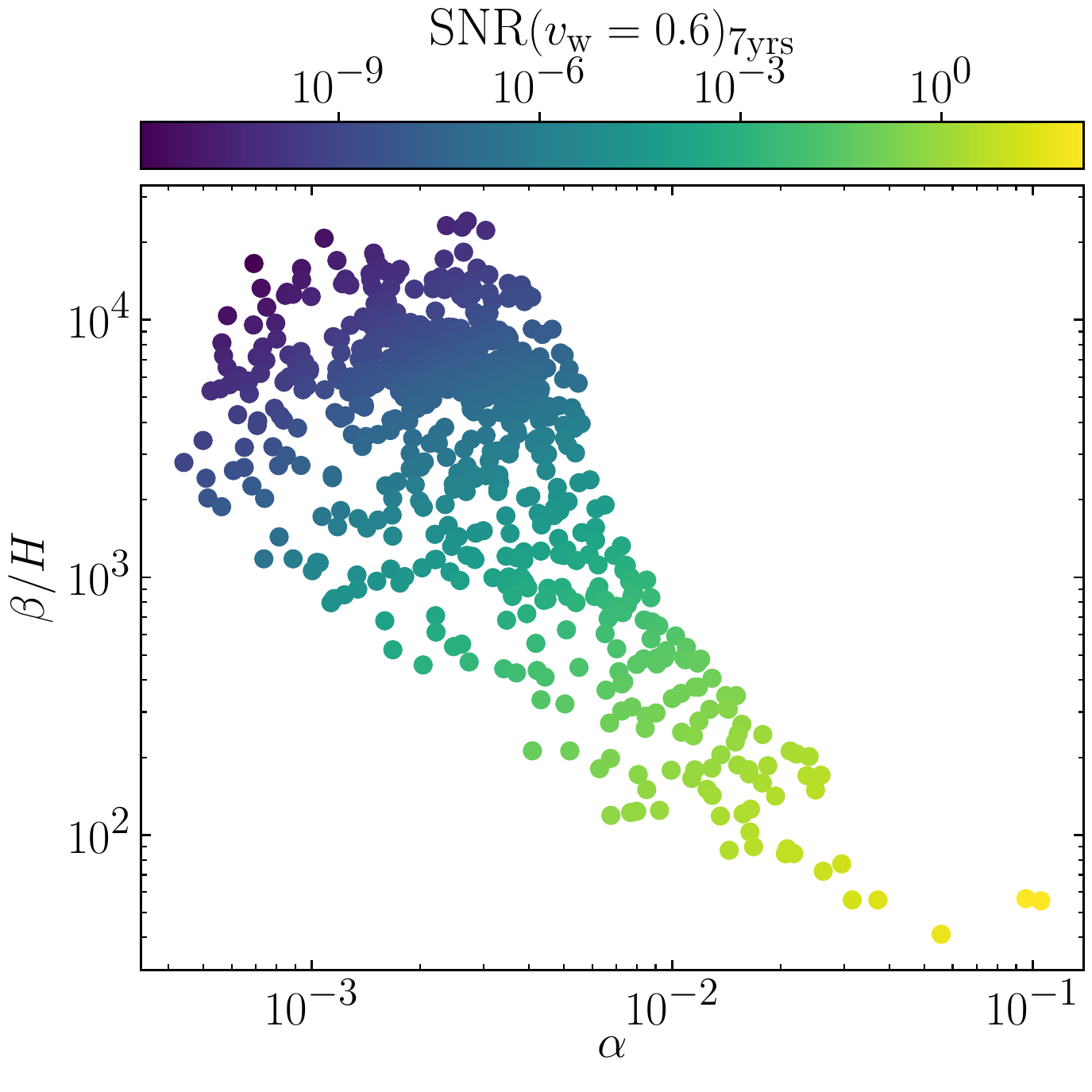}
\caption{\small Parameter points of the scan shown in~\reffi{Fig1} in the $(\alpha,\,\beta/H)$ plane, with the color-code indicating the SNR at LISA 
(assuming $v_{\rm w}=0.6$ and $\mathcal{T} = 7$ years).}
\label{fig4}
\end{figure}

In \reffi{fig4} we show the values of the inverse duration of the phase
transition $\beta/H$ in dependence of the strength $\alpha$ for all the points in our random scan satisfying $\xi_{n}>1$ (region E in \reffi{fig2}). The color code indicates
the value of the SNR at LISA (for $v_{w}=0.6$ and a LISA mission duration $\mathcal{T} = 7$ years). As expected, the points with the largest values of $\alpha$ and the smallest values of $\beta/H$ feature the largest SNRs for LISA.
The SNR values range over several orders of magnitude for
relatively small changes in the values of the masses $m_{H}$ and $m_{A}$, as 
will be shown below.
This is a consequence of the strong sensitivity of the predicted GW spectra to the underlying 2HDM model parameters (specifically, the BSM scalar masses).\footnote{Such a strong sensitivity has already been observed in \citere{Dorsch:2016nrg} (see, for instance, Fig.~3 therein). Similar observations have been made in the triplet extension of the SM~\cite{Friedrich:2022cak}.}
We also note that 
\GW{within the parameter space displayed in \reffi{fig3}}
the strongest GW signals are concentrated in a very narrow region of the ($m_{H}$, $m_{A}$) mass plane adjacent to the parameter space featuring vacuum trapping, 
and thus only a very small fraction of the 2HDM neutral BSM mass plane could be probed by LISA.

\begin{figure}[t]
\centering
\includegraphics[width=0.6\textwidth]{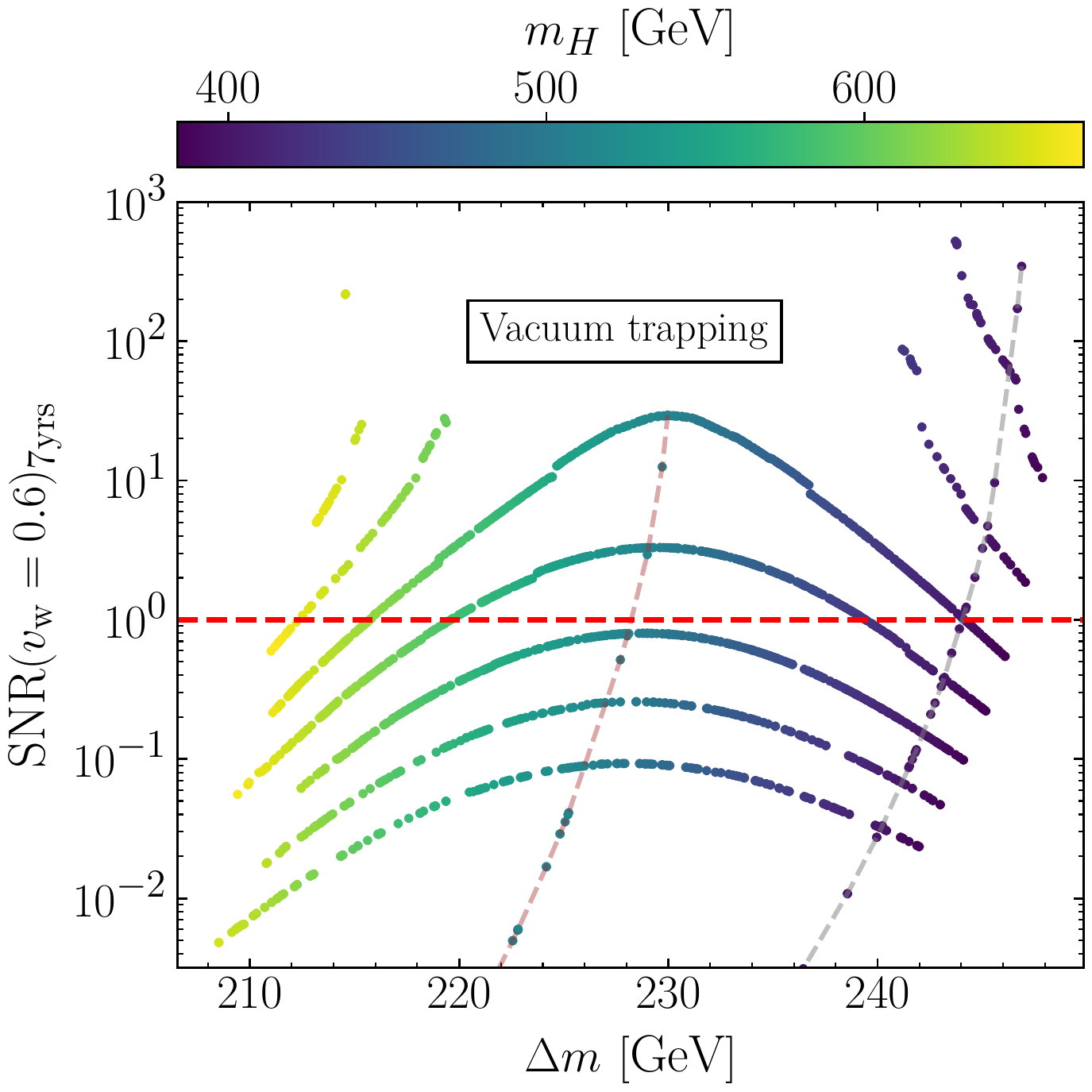}
\caption{\small SNR at LISA (for $v_w = 0.6$ and $\mathcal{T} = 7$ years) 
against $\Delta m = m_A - m_H$ for the parameter points of the dedicated finer scan (see text) with
$m_A = a\, m_H + b$, $a = 0.87$ and $b = \{291,292,293,294,295,296,297\} \gev$.
The color coding of the points indicates the values of $m_H$.
}
\label{fig5}
\end{figure}

In order to explore in detail the region of parameter space where the strongest GW signals are present, we have performed a linear regression of the points
featuring $\text{SNR} \gtrsim 0.5$, which are effectively found along a line 
given by $m_{A} = a\,m_{H} + b$, with $a=0.87$ and $b=295 \gev$. We have then performed a finer scan of the regions adjacent to this line
along parallel lines in the $m_H$-$m_A$ plane by shifting the 
value of $b$ in steps of $ 1 \gev$, i.e. for $b\in\left\{291,292,293,294,295,296,297\right\}$ GeV.
The results of this dedicated finer scan can be seen in \reffi{fig5}, where we show the GW SNR at LISA 
in dependence of the mass 
difference $\Delta m \equiv m_{A}-m_{H}$ (we recall that we set
$m_A = m_{H^\pm}$ and $M=m_H$ throughout this work).
The color code indicates the value of $m_{H}$. 
Bearing in mind the large uncertainties of the predictions for the
GW signal from a FOEWPT, as discussed in section \ref{sec:gws},
we consider as potentially detectable by LISA any SNR of $\mathcal{O}(1)$,
and mark the corresponding (indicative) threshold $\text{SNR}=1$ in 
\reffi{fig5} as a horizontal dashed red line.
The largest SNR values that we find in our finer scan are $\mathcal{O}(100)$ to $\mathcal{O}(1000)$ (such points could therefore be detected by LISA for $\mathcal{T} < 7$ years and/or with a substantially different assumption on $v_w$).
For $b = 296, \, 297$ GeV, \reffi{fig5} shows a region ranging from $\Delta m \sim 215 \gev$ to $\Delta m \sim 240 \gev$ where the 
\GW{parameter} points yielding the largest SNR values are found to be unphysical 
\GW{as a consequence of} vacuum trapping (the corresponding lines of \GW{benchmark points} in \reffi{fig5} are thus interrupted in this region).
Large values of the SNR 
\GW{of $\mathcal{O}(100\gev)$ or above}
are only found at the lower and the upper end of the $\Delta m$ scan range, where the 
\GW{occurrence of vacuum trapping is just barely avoided}. 
In fact, a further 
\GW{scanned} line of parameter points in \reffi{fig5} with $b = 298\gev$ is entirely excluded as a result of vacuum trapping. 

In addition to the finer scan discussed above, we show in \reffi{fig5} the SNR resulting from scans with fixed value of $m_{H}$ and increasing $\Delta m$, specifically for $m_{H} = 400 \gev$ (gray dashed line in \reffi{fig5}) and $m_H = 511 \gev$ (brown dashed line in \reffi{fig5}).
\htm{These additional lines
make even more visible the
drastic change of the SNR at LISA as a
consequence of a variation of the masses
$m_A = m_{H^\pm}$ by only a few~GeV.
Moreover, both} lines show the same
features regarding vacuum trapping
as discussed above.
This whole analysis 
demonstrates that the phenomenon of vacuum trapping severely limits the possibility of achieving large values of the SNR at LISA from GW production in the 2HDM.

\begin{figure}[t]
\centering
\includegraphics[width=0.48\textwidth]{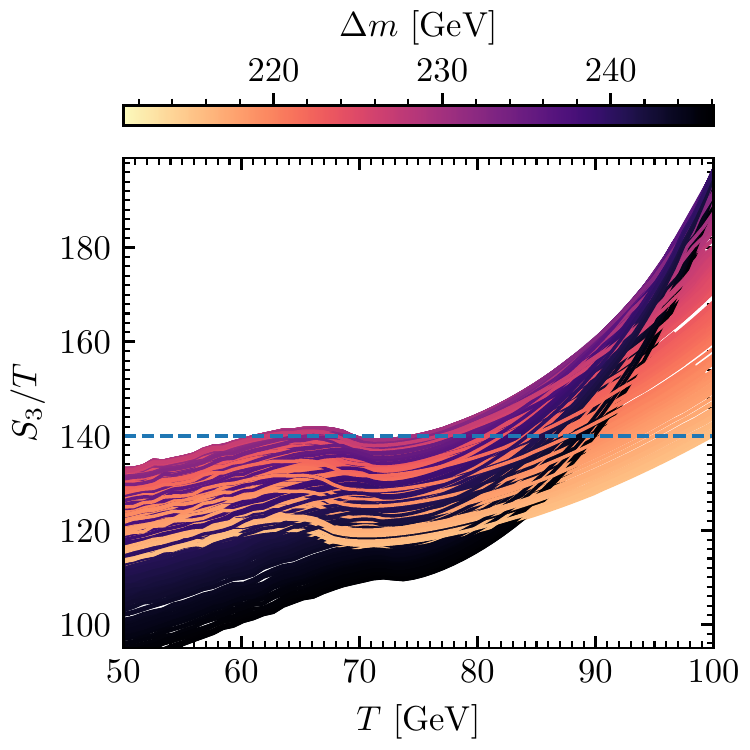}~
\includegraphics[width=0.48\textwidth]{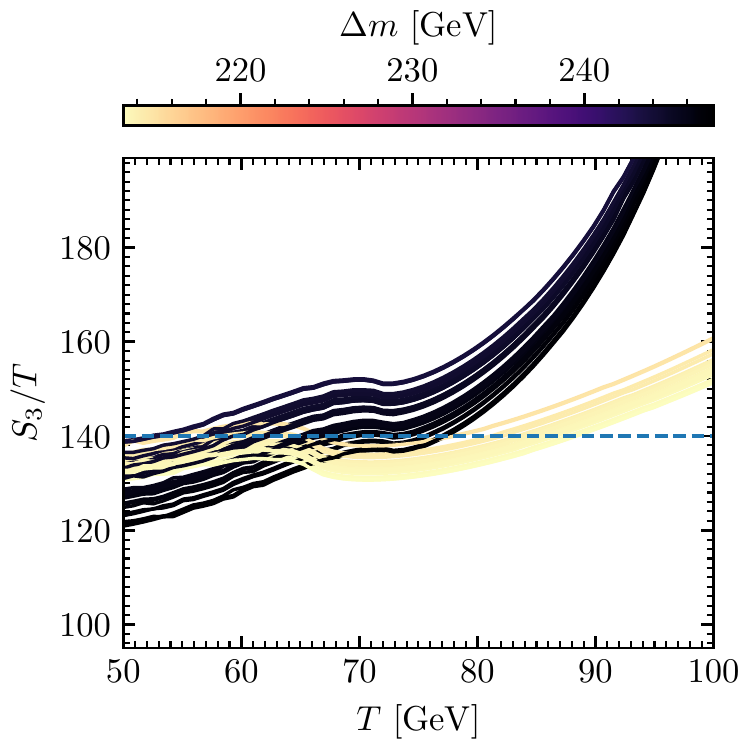}
\caption{\small
$S_3(T)/T$ as a function of $T$ with the color coding indicating the value of
$\Delta m = m_A - m_H$. In the left plot we show the results for scan points corresponding to $b = 295$ GeV in the dedicated scan of \protect\reffi{fig5},
whereas in the right plot we show the results for the $b = 297$ GeV line of points
(which is interrupted at intermediate values of $\Delta m$ due to the presence of vacuum trapping). The dashed blue horizontal line indicates $S_3(T)/T = 140$. The crossing of the lines for $S_3(T)/T$ with the dashed-blue line for decreasing $T$ signals
the onset of the phase transition at the respective temperature (see the nucleation criterion in \refeq{cond_trans}).}
\label{fig:actions}
\end{figure}

\vspace{2mm}

The strong dependence of the SNR on the 2HDM model parameters, pointed out at the beginning of this section and shown explicitly in \reffi{fig5}, is related to the fact that the largest 
GW signals occur 
just at the border of the parameter space region in which the Universe remains trapped in the false vacuum. In order to investigate this in more detail, we depict in \reffi{fig:actions} the values of the bounce action over the temperature, $S_3(T)/T$, for temperatures lower than $T_c$, such that a FOEWPT can occur. In the left panel of \reffi{fig:actions} we show $S_3(T)/T$ for $b = 295$ GeV in our detailed scan from \reffi{fig5} (corresponding to the benchmark line in \reffi{fig5} with the largest values of the SNR without featuring a gap as a consequence of vacuum trapping):    
bearing in mind that we assume \GW{that} the onset of the FOEWPT occurs for $S_3(T)/T \sim 140$ (recall the discussion in section~\ref{sec:FOEWPT}), we see that the benchmark 
\GW{lines in the left plot of} \reffi{fig:actions} with $\Delta m \sim 230$ GeV barely reach  $S_3(T)/T \sim 140$, and are thus on the verge of being vacuum-trapped.
In the right panel of \reffi{fig:actions} we show the corresponding values of 
$S_3(T)/T$ for the $b = 297$ GeV benchmark set, which features vacuum trapping for 
$\Delta m$ in the approximate range $[215,\, 240] \gev$ (as seen in \reffi{fig5}).
As a result, the lines in \GW{the right plot of} \reffi{fig:actions} are separated into two different bundles. \GW{For the parameter 
lines in between these two bundles the prediction remains} above $S_3(T)/T = 140$ (depicted as as dashed blue line) over the whole temperature interval $0 < T < T_c$, reflecting vacuum trapping \GW{(and those lines are therefore not depicted)}. In addition, many $S_3(T)/T$ lines have their minima just below the dashed blue line. Since they are on the verge of vacuum-trapping, these lines become rather flat as they approach $S_3(T)/T = 140$, leading to a large variation of $T_n$ (i.e.\ the temperature at which $S_3(T)/T \simeq 140$ is achieved) within a very small $\Delta m$ range. 
As an example,  for the black bundle of lines in \GW{the right plot of} \reffi{fig:actions} we have  $243\gev < \Delta m < 247\gev$, \GW{i.e.\ a variation within just $4\gev$, while} $T_n$ varies in the range $52\gev < T_n < 77\gev$. At the same time, by comparing the two panels of \reffi{fig:actions} we observe that a very small change in $b$ from our detailed scan leads to large variations of the $T_n$ behaviour as a function of $\Delta m$.
The very strong dependence\footnote{We stress that the FOEWPT nucleation criterion used here, $S_3(T)/T = 140$, is only an approximation~\cite{Caprini:2019egz}, and also the computation of the tunneling rate given by \refeq{eqfuncdeter} suffers from sizable theoretical uncertainties from missing higher-order contributions (both in the prefactor $A(T)$ and in the perturbative formulation of $V_{\rm eff}$, affecting $S_3$) as well as from the issues of gauge dependence~\cite{Patel:2011th}
and renormalization scale
dependence (see \refapp{app:scale}).
Yet, such uncertainties only have a sizable impact on parameter points close to the vacuum-trapping region, whereas regions leading to weaker GW signals 
(i.e.\ \GW{regions that are} not in the vicinity of the vacuum-trapping region) do not feature such large uncertainties in the SNR prediction. Thus, our conclusion that 
\GW{within the parameter region featuring a FOEWPT the points giving 
rise to a GW signal that could potentially be observable at LISA occur only in a small 
region in the vicinity of the region that is excluded by vacuum trapping}
is therefore robust even in view of these issues.} 
of $T_n$ on subtle changes of the 2HDM masses then feeds into the GW spectra (e.g.\ $\alpha \sim 1 / T_n^4$) and ultimately into the SNRs at LISA. 
As a result, values of the $\mathrm{SNR} > 1$ are found only in a very restricted region of the 2HDM parameter space, in the vicinity of the vacuum-trapping (unphysical) parameter region.

\begin{figure}
\centering
\includegraphics[width=0.48\textwidth]{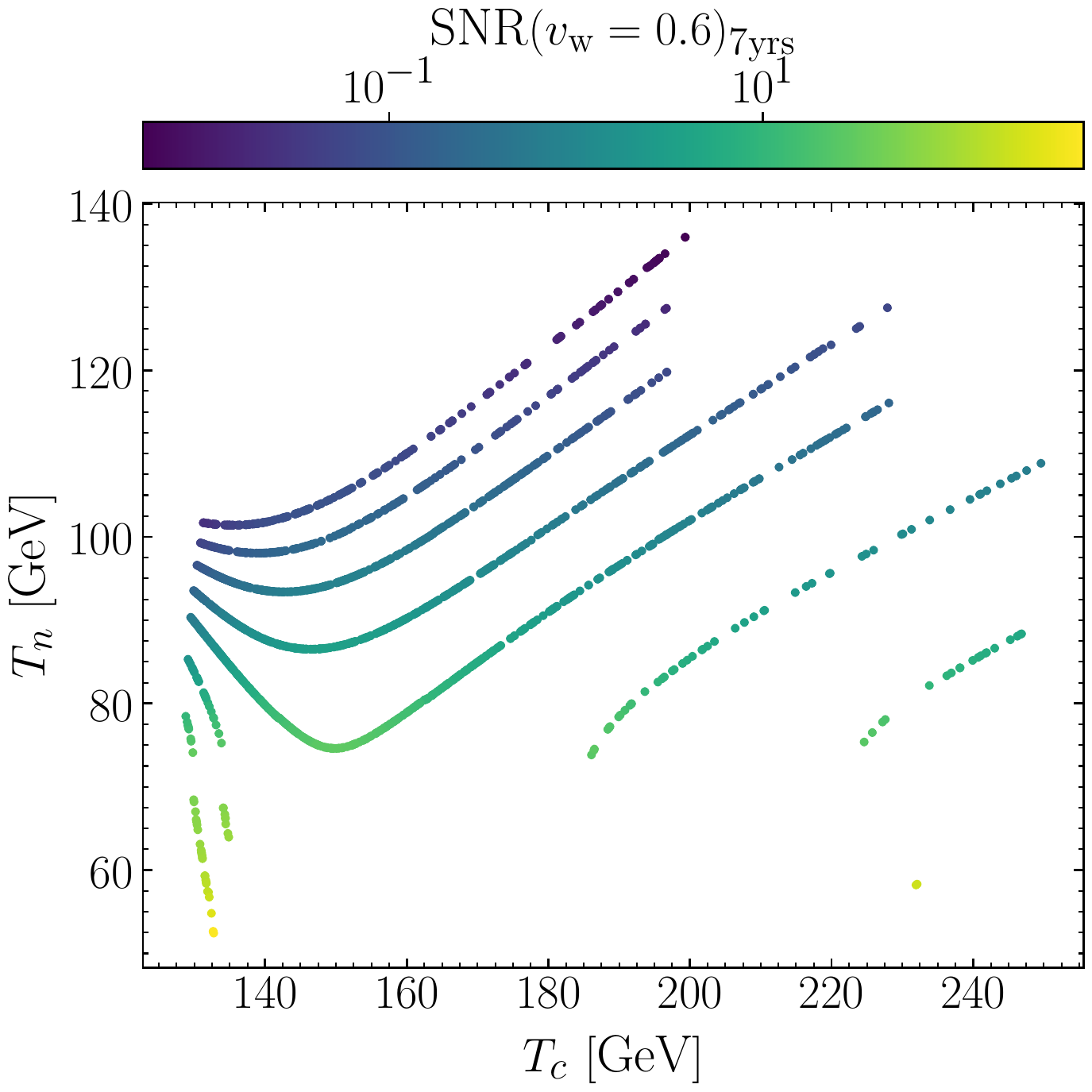}~
\includegraphics[width=0.48\textwidth]{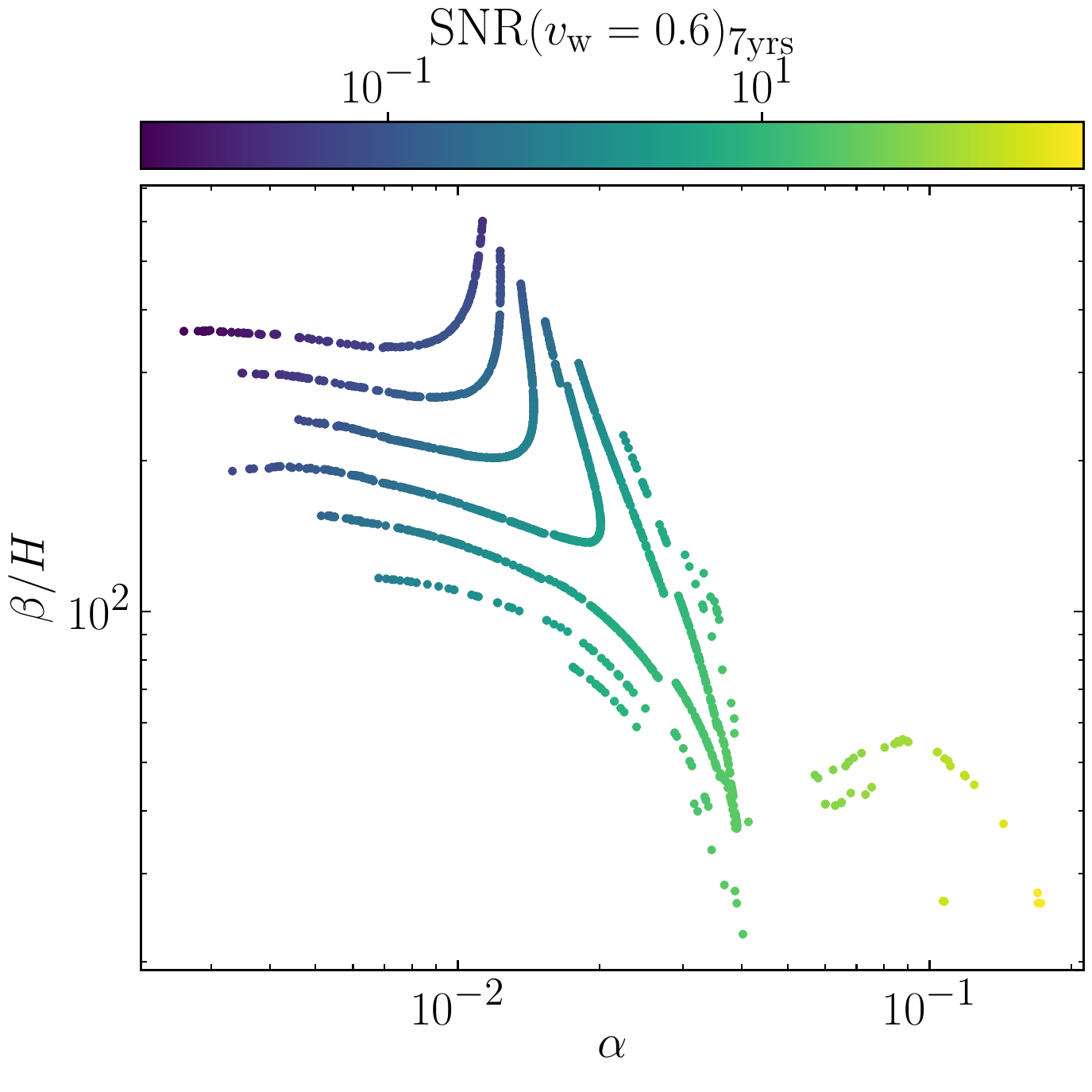}
\caption{\small Parameter points of the dedicated finer scan of \reffi{fig5}, 
in the $(T_c,T_n)$ plane (left panel)
and in the $(\alpha,\, \beta / H)$ plane (right panel), with the color coding of the points indicating the SNR at LISA.}
\label{fig6}
\end{figure}

\vspace{2mm}

In \reffi{fig6} we explicitly show, for the detailed scan introduced in \reffi{fig5}, the dependence of the LISA SNR on the quantities $T_{n}$, $\alpha$ and $\beta/H$.
In \GW{the left plot of} \reffi{fig6} we show the relation between the nucleation temperature $T_n$ and the critical temperature $T_c$ for this scan (with color-code indicating the SNR at LISA). The large difference between \GW{the two} temperatures for all the points in this scan reaffirms the necessity of computing the nucleation temperature in order to make reliable predictions concerning the FOEWPT properties in the 2HDM, since not even a qualitative description of the strength of the phase transition is possible based on the knowledge of the critical temperature.  
In the right panel of \reffi{fig6} we show the corresponding detailed scan points in the ($\alpha$, $\beta/H$) plane, from which an intricate dependence of both parameters on the 2HDM masses can be inferred by correlating with the information from \reffi{fig5}. 
Compared to the broader scan shown in \reffi{fig4}, we find here a substantially smaller range of $\beta / H$ (down to $\beta/H \sim 23$) and overall larger values of $\alpha$ (up to $\alpha \sim 0.17$).
We stress here that values of $\beta / H \ll 100$ are an indicator of \GW{a scenario that is} close to featuring vacuum trapping (see e.g.\ the discussion in~\cite{Ellis:2018mja}).

\vspace{2mm}

\begin{figure}[t]
\centering
\includegraphics[width=0.6\textwidth]{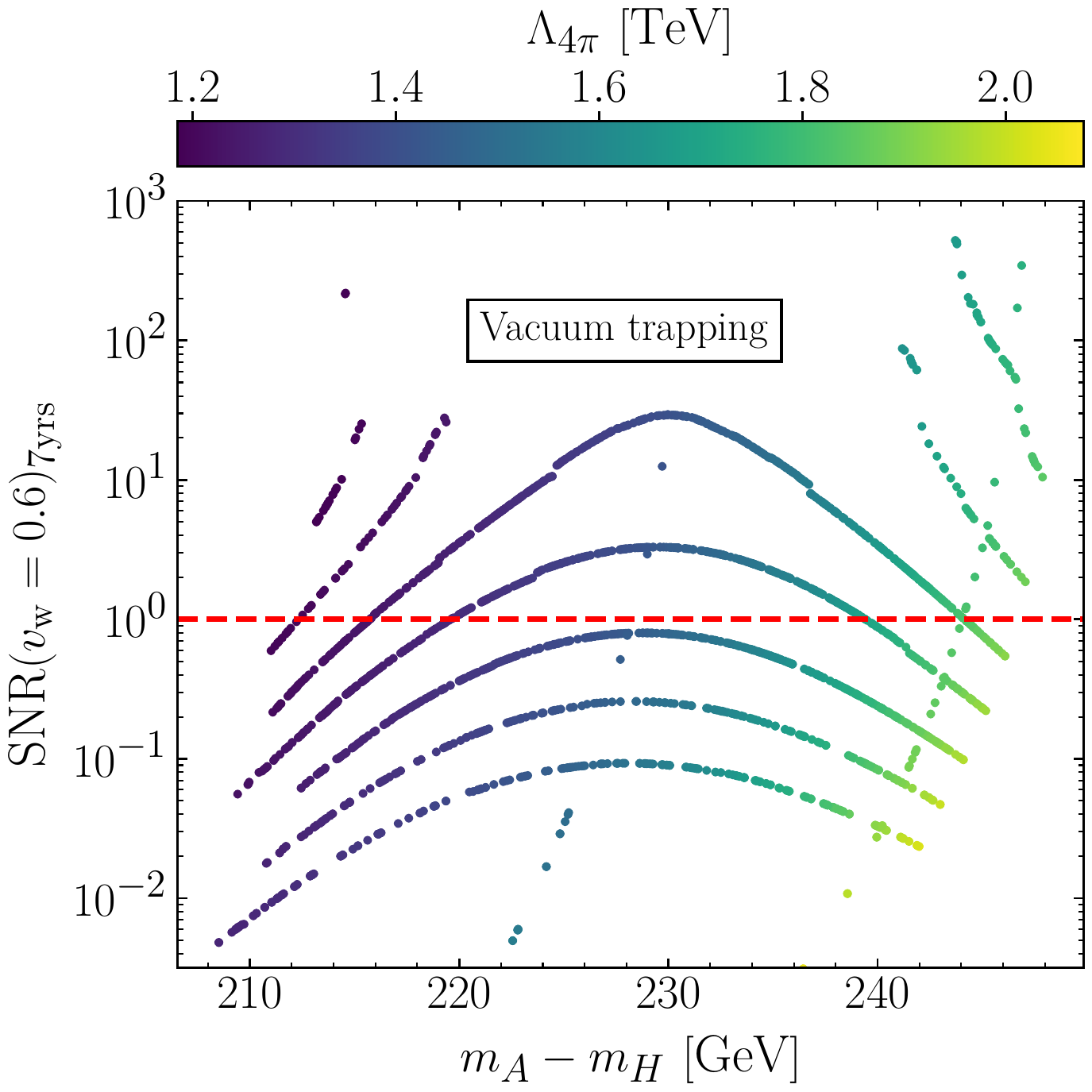}
\caption{\small
Parameter points of the dedicated finer scan of \reffi{fig5}, 
in the ${(\Delta m = m_A - m_H,\mathrm{SNR})}$ plane, with the
color coding indicating the energy scale $\Lambda_{4 \pi}$ at
which one of the quartic couplings reaches the naive perturbative bound $4\pi$.}
\label{fig7}
\end{figure}

Finally, we 
\GW{emphasize once more} that a FOEWPT in the 2HDM requires sizable quartic scalar couplings $\lambda_i$ \GW{such that} a potential barrier between the two minima involved in the transition \GW{can} be generated via radiative and/or thermal loop corrections. 
The RGE evolution of such sizable quartic couplings can drive the theory into a non-perturbative regime already at energies not far from the TeV~scale, as discussed in detail in section \ref{sec:rges} (see also \citere{Basler:2017nzu} for a one-loop analysis). This issue is most severe for the strongest phase transitions, such as the ones that produce GW signals with sizable SNR values at LISA. We 
\GW{therefore}
investigate the energy range in which the theory is well-defined for the type~II 2HDM parameter regions that could yield an observable GW signal at LISA.
In \reffi{fig7} we show the 2HDM parameter points of our detailed scan in the  $(\Delta m = m_A - m_H,\,\mathrm{SNR})$ plane, as in \reffi{fig5}, but now with the color-coding indicating the energy scale $\Lambda_{4 \pi}$ at which one of the quartic scalar couplings $\lambda_i$ reaches the naive perturbative bound $4\pi$ (see section \ref{sec:rges} for details).
The value of $\Lambda_{4 \pi}$ signals the energy scale $\mu$ at (or below) which new BSM physics should be present in order to avoid a Landau pole and \GW{to} render the theory well-behaved above that energy scale.
We observe that the lowest values of $\Lambda_{4 \pi}$ in our detailed scan are $\Lambda_{4 \pi}\sim 1.2 \tev$, whereas the largest values are found slightly above $\Lambda_{4 \pi} = 2 \tev$. By comparing with \reffi{fig5} we also observe that the smallest values of $\Lambda_{4 \pi}$ correlate with the largest values of $m_H$ in the scan, which can have important phenomenological implications (as we discuss in the next section). Altogether, \reffi{fig7} shows that parameter regions that feature a potentially detectable (SNR $> 1$) GW signal at LISA would require new-physics effects (e.g.\ new strongly coupled states) at energy scales that are well within the reach of the LHC. 
\GW{This finding calls for a thorough} assessment of the complementarity between LHC (and future collider) searches and GW probes with LISA in these theories.

\subsection{Interplay between the \GW{HL-LHC (and beyond)} and LISA}
\label{sec:LISA_colliders}

As already outlined above, the 2HDM parameter regions featuring a GW signal \GW{that could potentially be} observable at LISA generally predict signatures of BSM physics within reach of the LHC, both from the presence of the 2HDM scalars themselves and from further new (strongly coupled) states that would \GW{in some parts of the parameter space} 
be needed to prevent the appearance of a Landau pole close to the TeV scale. In this section, we focus on the collider signals of the 2HDM scalars
\GW{in view of the prospects for}
the interplay between the possible observation of a stochastic GW signal from the 2HDM at LISA and 
\GW{collider probes (at the HL-LHC and a future $e^+e^-$ Linear Collider)}
of the 2HDM states.

\subsubsection{GWs at LISA vs.~direct BSM searches at \GW{the} LHC}
\label{sec:gw-dd}

\begin{figure}[t]
\centering
\includegraphics[width=0.61\textwidth]{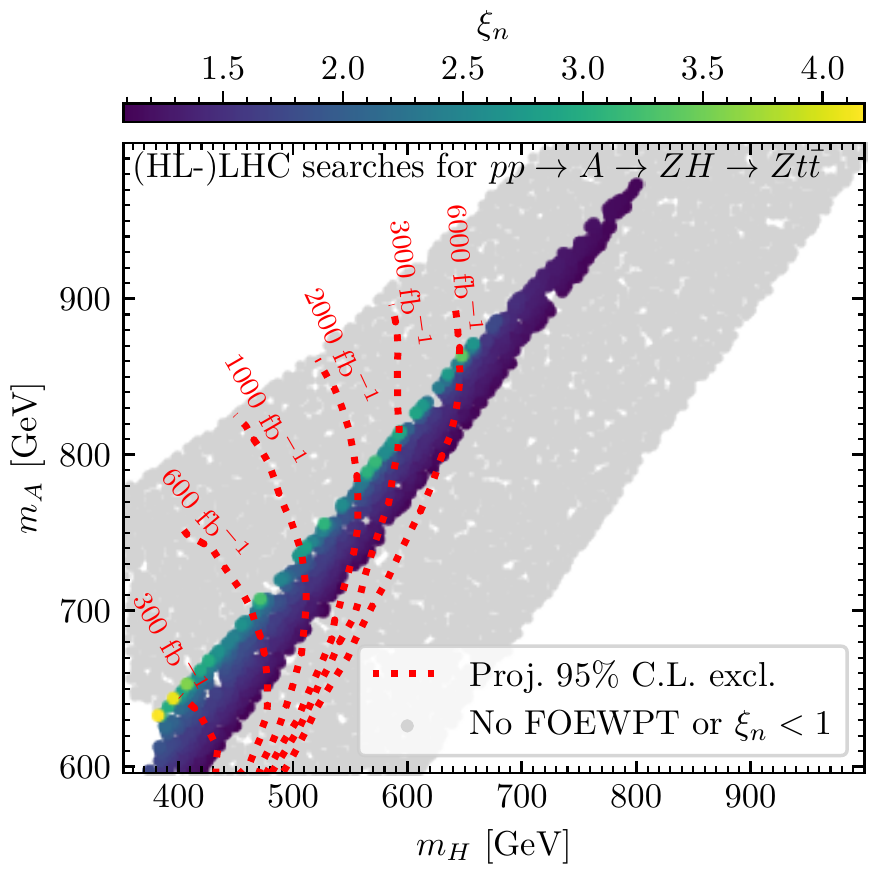}
\caption{\small
Parameter points of the 
scan discussed in 
\refse{sec:2hdm_thhist} in the $(m_H,m_A)$ plane, with the color coding indicating the value of $\xi_n$ for the points that feature a strong FOEWPT, i.e.~$\xi_n > 1$. The remaining points are shown in gray. The red dashed lines indicate the projected 95\% C.L.~exclusion regions resulting from the (HL-)LHC searches for the process $pp \to A \to ZH$ with $H$ decaying into a pair of top quarks (see text for details).}
\label{fig7_lhc}
\end{figure}

Given the projected HL-LHC and LISA timelines, the HL-LHC 
\GW{is expected to} scrutinize the 2HDM parameter space of relevance for GW searches before the LISA observatory will start taking data. We show that, within the type~II 2HDM,
\GW{for the case where no}
direct BSM signatures \GW{will be detected} at the high-luminosity phase of the LHC \GW{the resulting limits would essentially exclude the parameter regions 
giving rise to a potentially observable GW signal at LISA. Thus, 
the prospects for observing a GW signal at LISA crucially depend on the outcome of the high-luminosity phase of the LHC.}

Among the possible collider signatures of the heavy 2HDM scalars, the most promising ones to probe the 2HDM parameter \GW{region featuring} a FOEWPT consist of Higgs
\htm{boson} cascade decays, due to the sizable mass splittings between the BSM Higgs bosons. Specifically, the production of the 
\GW{CP-odd Higgs boson} $A$ that then decays into a $Z$ boson and the heavy CP-even scalar $H$ is a \textit{smoking-gun} collider signature of FOEWPT scenarios in the 2HDM~\cite{Dorsch:2014qja}. 
This signature has been searched for at the LHC with $\sqrt{s} = 8\tev$ and $13\tev$ assuming that $A$ is produced via gluon-fusion or in association with a pair of bottom quarks, and utilizing the leptonic decay modes of the $Z$-boson. The scalar $H$ was required to decay either to a pair of bottom quarks or to a pair of tau leptons~\cite{CMS:2016xnc,ATLAS:2018oht,CMS:2019ogx}. However, as already pointed out in \citere{Biekotter:2021ysx}, the combination of
\GW{the current}
theoretical and experimental constraints in the type~II 2HDM 
pushes $m_H$ to be above the di-top threshold in almost the entire parameter region featuring a FOEWPT. 
Then, the branching fractions for $H \to b\bar{b}$ and $H \to \tau^+\tau^-$ become very small (except for large values of $t_\beta$), and searches via these final states do not yield relevant constraints on FOEWPT scenarios.
It is instead much more promising to search for $A \to Z H$ signatures with $H$ decaying into a pair of top quarks, and preliminary studies of this final state \GW{exist} 
in the literature~\cite{Haisch:2018djm,Goncalves:2022wbp}. 
\GW{Efforts} to analyze the $Z\, t \bar t$ final state are ongoing by both the ATLAS~\cite{dpgatlas} and CMS~\cite{dpgcms,fischerthesis} collaborations. 
We here use the public 
results on this channel
obtained in a Master thesis for the CMS Collaboration
(using only the $Z \to \mu^+ \mu^-$ decay mode) for an integrated luminosity of $41\, \mathrm{fb}^{-1}$ at $13\tev$~\cite{dpgcms} to estimate the projected (HL-)LHC sensitivity to the process $A \to ZH$ in the $Z\, t \bar t$ final state for several integrated luminosities: $\mathcal{L} = 300\, \mathrm{fb}^{-1},\ 600\, \mathrm{fb}^{-1},\ 1000\, \mathrm{fb}^{-1},\ 2000\, \mathrm{fb}^{-1},\
3000\, \mathrm{fb}^{-1}$ and $6000\, \mathrm{fb}^{-1}$ (the latter corresponds to the \GW{expected combined} total integrated luminosity \GW{that will be} collected by ATLAS and CMS at the HL-LHC). 
We obtain the predicted 2HDM production cross sections (at NNLO) times branching ratios for the $p p \to A \to Z H \to \mu^+ \mu^- \, t \bar t$ signature as a function of $m_A$ and $m_H$ (with the rest of parameters fixed as in~\refeq{eqranges1}) using \texttt{SusHi}~\cite{Harlander:2012pb} and \texttt{N2HDECAY}~\cite{Engeln:2018mbg}. 
In \reffi{fig7_lhc} we show the expected 95\% C.L.\ exclusion sensitivity for different values of $\mathcal{L}$ from a naive rescaling of the CMS expected limits by a factor $\sqrt{(41\, \mathrm{fb}^{-1}) / \mathcal{L}}$ (which assumes that the present CMS sensitivity is limited by statistics rather than systematics).
We emphasize that taking into account also other (leptonic) decay modes of the $Z$ boson yields even stronger projected limits~\cite{fischerthesis}.
On the other hand, the preliminary
projected cross-section
limits do not yet account for all
systematic uncertainties, for instance,
from the $b$-tagging efficiencies.
The inclusion of such
systematic uncertainties 
\GW{could weaken}
the expected
sensitivity. Considering both aspects, 
the exclusion regions shown in \reffi{fig7_lhc} 
can be regarded as fairly conservative estimates.
Nevertheless, we verified that even assuming
that the cross-section
limits are a factor of~2 weaker,
the HL-LHC could exclude the parameter region
featuring a strong FOEWPT up to masses of
$m_H \sim 550\gev$ and $m_A \sim 750\gev$.

\GW{The projected exclusion limits in
\reffi{fig7_lhc} are compared with} the points of the 2HDM parameter scan discussed in section \ref{sec:2hdm_thhist}, where the parameter points featuring a strong FOEWPT are shown in color (the color-coding indicates the value of $\xi_n$), and the remaining points are depicted in gray. 
Already at the end of LHC Run~3 with $300\, \mathrm{fb}^{-1}$ ($600\, \mathrm{fb}^{-1}$ assuming a potential combination of ATLAS and CMS~data), a substantial part of the parameter space 
\GW{featuring a strong FOEWPT}
will be explored, \GW{corresponding to
values of} $m_H \lesssim 470$~GeV
\htm{(see \reffi{fig5})}. In particular,
the 2HDM region yielding observable GW
signals at LISA with values of
$\Lambda_{4\pi} > 2$~TeV \htm{(see \reffi{fig7})}
will be completely covered by this LHC search during Run~3, and so will be the parameter points with the strongest phase transitions, corresponding to values of $\xi_n \sim 4$. 
The HL-LHC, with ten times more data, will be able to probe masses up to $m_H \sim 650\gev$ via the $A \to ZH$ ($H \to t \bar t$) search, covering almost the entire 2HDM region that features a GW signal \GW{that could potentially be} detectable with LISA (see \reffi{fig5}).
This analysis highlights the importance of putting the expectations for GW signals from FOEWPTs that could be detectable by LISA into the context of the projected (HL-)LHC results.

\subsubsection{GWs at LISA vs.~Higgs boson self-coupling measurements at LHC and ILC}
\label{sec:gw-lahhh}

\begin{figure}[t]
\centering
\includegraphics[width=0.48\textwidth]{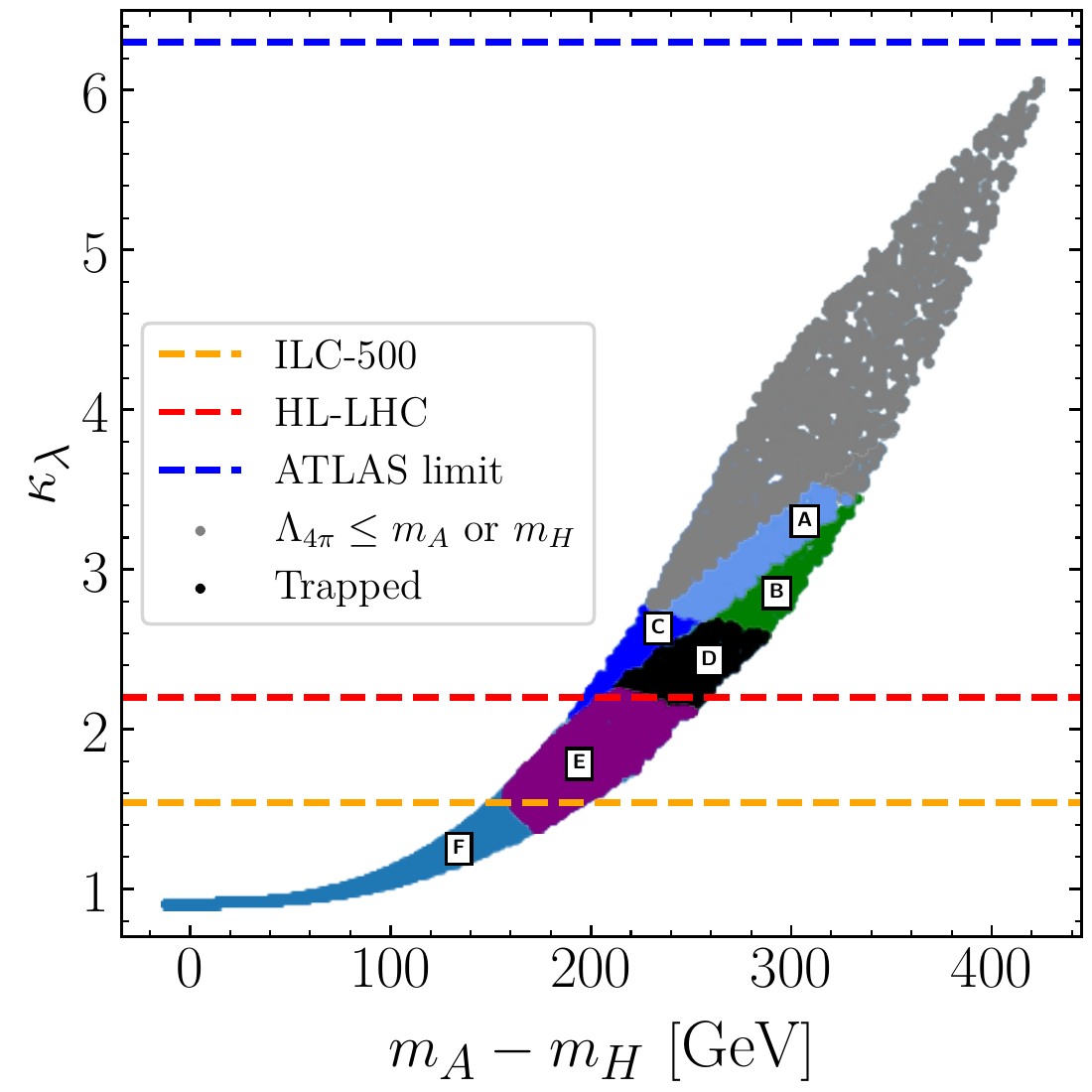}~
\includegraphics[width=0.48\textwidth]{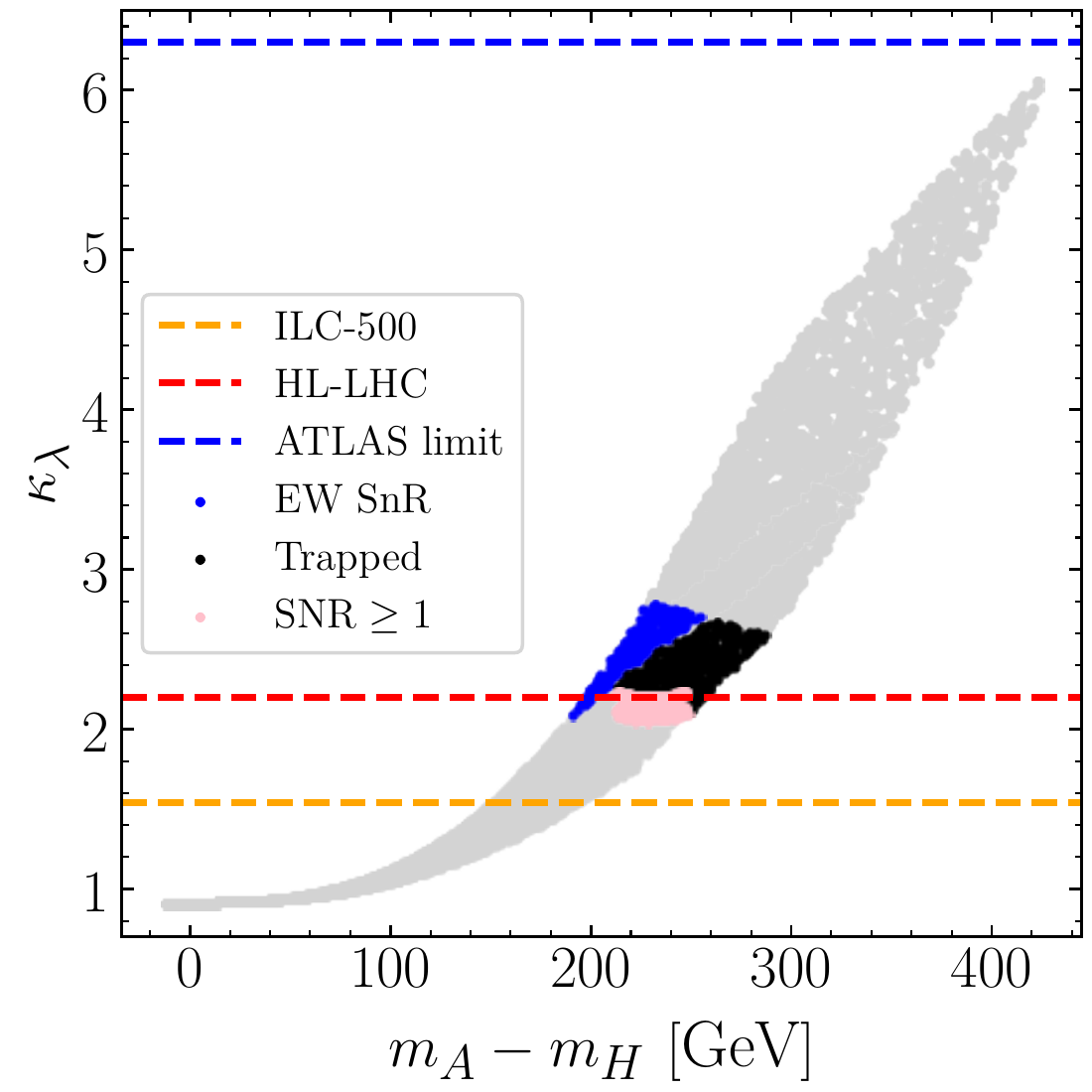}

\caption{\small Parameter points from the scan as defined in \refeq{eqranges1} with the mass difference $m_A - m_H$ on the horizontal axis and $\kappa_\lambda$ on the vertical axis. In the left panel, the color of the points indicates the different kinds of thermal histories: the letter specifies each region according to \reffi{fig2}, and dark-gray points feature $\Lambda_{4 \pi} \leq m_A$ or $m_H$, and/or a short-lived EW vacuum. In the right panel, blue points feature EW SnR, black points feature vacuum trapping (and are therefore unphysical), and pink points predict a FOEWPT with an associated GW signal that could be detectable at LISA ($\mathrm{SNR} \geq 1$, see text for details). The characteristics of the light-gray points can be inferred from the left panel.
\GW{Also shown in both plots is the current limit on $\kappa_\lambda$ from ATLAS as well as the projected sensitivities of the HL-LHC and the ILC running at a c.m.\ energy of $500\gev$.}}
\label{fig8}
\end{figure}

A well-known avenue to probe the thermal history of the EW symmetry, particularly in connection with a possible FOEWPT, is the measurement of the trilinear self-coupling of the \GW{Higgs boson at} $125\gev$. \GW{A FOEWPT is} generically associated with a sizable enhancement of the trilinear self-coupling $\lambda_{hhh}$ as compared to the SM prediction~\cite{Noble:2007kk,Huang:2015tdv}.\footnote{This is especially the case for FOEWPTs which are not \textit{singlet-driven} (caused by a singlet scalar field coupling to the \GW{SM-like} Higgs doublet). For a singlet-driven FOEWPT, it is possible to avoid such large enhancements~\cite{Ashoorioon:2009nf}.} 
In the following, we determine the values of $\lambda_{hhh}$ predicted in the 2HDM parameter space regions which feature a FOEWPT, including the regions that would yield a GW signal \GW{that could} potentially be observable at LISA. 
According to our \GW{treatment} of the zero-temperature effective potential from \refeq{Veff_Zero}, $\lambda_{hhh}$ is calculated here at the one-loop level 
\GW{(see \citere{Bahl:2022jnx} for a discussion of the impact of the dominant two-loop contributions in the 2HDM). In order}
to align our analysis with the experimental interpretations of 
\GW{bounds on} the Higgs
\htm{boson} trilinear self-coupling 
\GW{obtained} by the ATLAS and CMS collaborations within the $\kappa$-framework, we here define $\kappa_\lambda = \lambda_{hhh} / \lambda_{hhh}^{\rm SM}$, where $\lambda_{hhh}^{\rm SM}$ is the tree-level Higgs \htm{boson} self-coupling prediction of the SM.
In \reffi{fig8} we show the \GW{predicted} values of $\kappa_\lambda$ in dependence of the mass splitting
$m_A - m_H$ for the parameter scan from \refeq{eqranges1}.
In the left panel, the various colors indicate the different types of thermal histories
(the letter in each region specifies the corresponding thermal evolution of the vacuum
according to the description of \reffi{fig2}).
As expected, large values of $m_A - m_H$ are correlated with large values of $\kappa_\lambda$.
In particular, parameter points featuring a strong FOEWPT (region~E) predict values of up to $\kappa_\lambda \sim 2$, and vacuum trapping (region~D) excludes part of the parameter space with even larger values of $\kappa_\lambda$. There are still 
physically viable parameter points predicting values of $\kappa_\lambda > 2$ (regions~A and~C; we remind the reader that region~B is unphysical, see section~\ref{sec:2hdm_thhist}), associated with the phenomenon of EW SnR. 
\GW{The plot shows that} the largest values of $\kappa_\lambda$ occur for 2HDM parameter regions that are not  phenomenologically viable (dark-gray points), as these regions feature an energy cutoff $\Lambda_{4 \pi}$ \GW{that is} smaller than the masses of the BSM scalar states, i.e.~$\Lambda_{4 \pi} < m_A = m_{H^\pm}$ or $\Lambda_{4 \pi} < m_H$; a large fraction of these points also features a short-lived EW vacuum (see \reffi{Fig1}).

The value of $\kappa_\lambda$ can be \GW{experimentally} constrained 
\GW{via information from}
double Higgs boson production at colliders
\GW{(and indirectly via measurements of single Higgs boson production)}. In order to compare the 2HDM predictions for $\kappa_\lambda$ with present and \GW{projected} future experimental constraints, we show in \reffi{fig8} the currently strongest 95\% C.L. experimental limit on $\kappa_\lambda$, corresponding to $\kappa_\lambda < 6.3$
as reported by ATLAS\footnote{CMS has reported a comparable upper limit of
$\kappa_\lambda < 6.49$~\cite{CMS:2022dwd}.} using the full Run~II dataset and combining measurements of single Higgs boson and (non-resonant) Higgs boson pair production~\cite{ATLAS-CONF-2022-050}. We also show the projected 95\% C.L.\ sensitivity of the HL-LHC (dashed red line), given by $\kappa_\lambda < 2.2$~\cite{Cepeda:2019klc}, and the projected 95\% C.L.\ sensitivity of the future International Linear Collider (ILC) with $\sqrt{s} = 500$ GeV and an integrated luminosity of $4000\, \mathrm{fb}^{-1}$ (dashed \GW{orange} line), given by $\kappa_\lambda < 1.54$~\cite{Durig:2016jrs}.
We stress that \GW{the current} experimental limits on $\kappa_\lambda$ hold under the assumption that the couplings of $h$ to other SM particles are those of the SM, which is the case in the alignment limit of the 2HDM (at leading order) used in this work. In addition, the projected limits shown for the HL-LHC and the ILC assume that $\kappa_\lambda = 1$ will be measured experimentally (we discuss below the 
\GW{consequences if a different value of $\kappa_\lambda$ is realized in nature}).\footnote{It should be noted that, with our definition of $\kappa_\lambda$ (which matches that of the ATLAS and CMS experimental collaborations), $\kappa_\lambda = 1$ corresponds to the SM prediction only \GW{if} one-loop corrections to $\lambda_{hhh}$ in the SM (which amount to $-9\%$ of the tree-level value~\cite{Dorsch:2017nza}) are neglected.} While the current experimental sensitivity \GW{at the LHC} is not sufficient to probe the viable parameter space analyzed here, the HL-LHC 
\GW{is expected to} be capable of 
\GW{probing essentially} the entire parameter space featuring EW SnR, \GW{and the} ILC-500 would furthermore probe most of the region featuring a strong FOEWPT, \GW{in particular the entire region with a GW signal that could be detectable at LISA (see below)}.

In order to estimate the values of $\kappa_\lambda$ for parameter points with detectable GW signals at LISA, we show in the right panel of \reffi{fig8} the same parameter plane as in the 
left panel, but with the strong FOEWPT parameter points predicting $\mathrm{SNR} \geq 1$ at LISA highlighted in light-pink. These points have values of $\kappa_\lambda \sim 2$, and thus lie 
\GW{near} the expected HL-LHC upper limit on $\kappa_\lambda$ \GW{and within the reach of the ILC running at $500\gev$.} 

\begin{figure}
\centering
\includegraphics[width=0.6\textwidth]{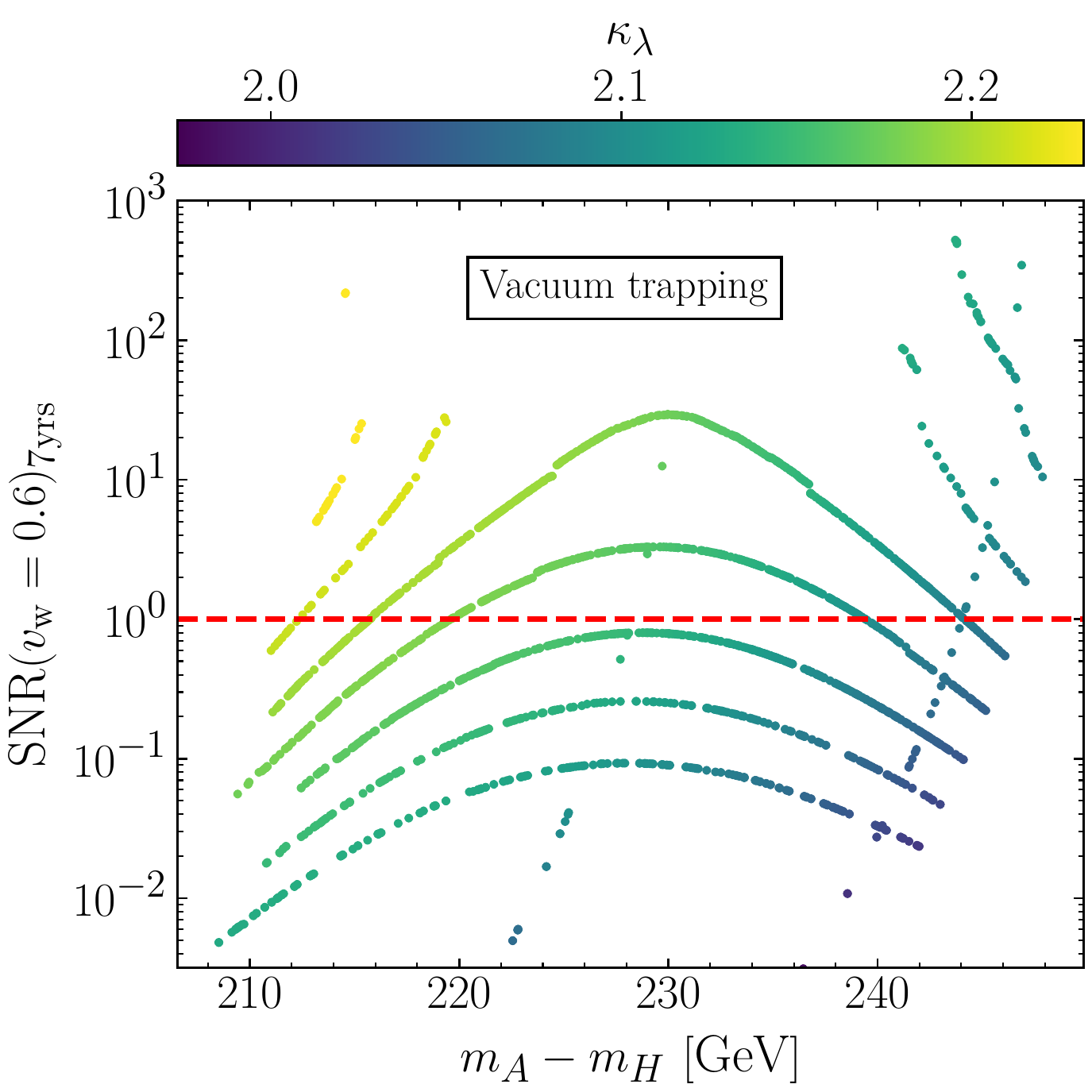}
\caption{\small Parameter points of the detailed finer scan discussed in section  \ref{sec:gravitationalwaves} (\GW{as} shown in \reffi{fig5} and \reffi{fig7}),
in the ${(\Delta m = m_A - m_H,\, \mathrm{SNR})}$ plane. The
color-coding here indicates the prediction for $\kappa_{\lambda}$.
}
\label{fig7b}
\end{figure}

To further scrutinize this parameter region, focusing on the interplay between measurements of the Higgs boson self-coupling at colliders and potential observations of GWs at LISA, we show in
\reffi{fig7b} the same plane as depicted in \reffi{fig5} and \reffi{fig7}, with the color-coding now indicating the values of $\kappa_\lambda$ (points above the dashed red line in \reffi{fig7b} 
\GW{therefore}
correspond to the pink area in \GW{the right plot of} \reffi{fig8}). \GW{The predicted values} of $\kappa_\lambda$ in this plot range from $\kappa_\lambda \sim 2$ up to $\kappa_\lambda \sim 2.2$, possibly within reach of the HL-LHC. The plot furthermore illustrates that a strong FOEWPT that gives rise to a potentially detectable GW signal is associated with a significant deviation from $\kappa_\lambda =1$ (see also \citere{Goncalves:2021egx}). 
Conversely, if no deviations of $\kappa_\lambda$ from the SM prediction are observed at the HL-LHC {and / or a future $e^+e^-$ Linear Collider running at $500\gev$}, no GW signal at LISA would be expected \GW{in the considered scenarios}. 

We also stress that future measurements of $\kappa_\lambda$ at the HL-LHC and the ILC will be
a very important probe of the EW phase transition, independently of the associated GW production (as shown in \reffi{fig8}, a large fraction of the parameter space featuring a strong FOEWPT does not yield an observable GW signal at LISA). 
\GW{We note in this context} that the leading two-loop corrections to the self-coupling of the SM-like Higgs boson can yield a sizable enhancement of $\kappa_\lambda$~\cite{Bahl:2022jnx}  with respect to the one-loop result. 
Thus, an analysis of $\kappa_\lambda$ at the two-loop level may result in even better prospects for a measurement of a modification of the Higgs boson self-coupling with respect to the SM value. \GW{We leave such} a study for the future.

\begin{figure}[t]
\centering
\includegraphics[width=0.98\textwidth]{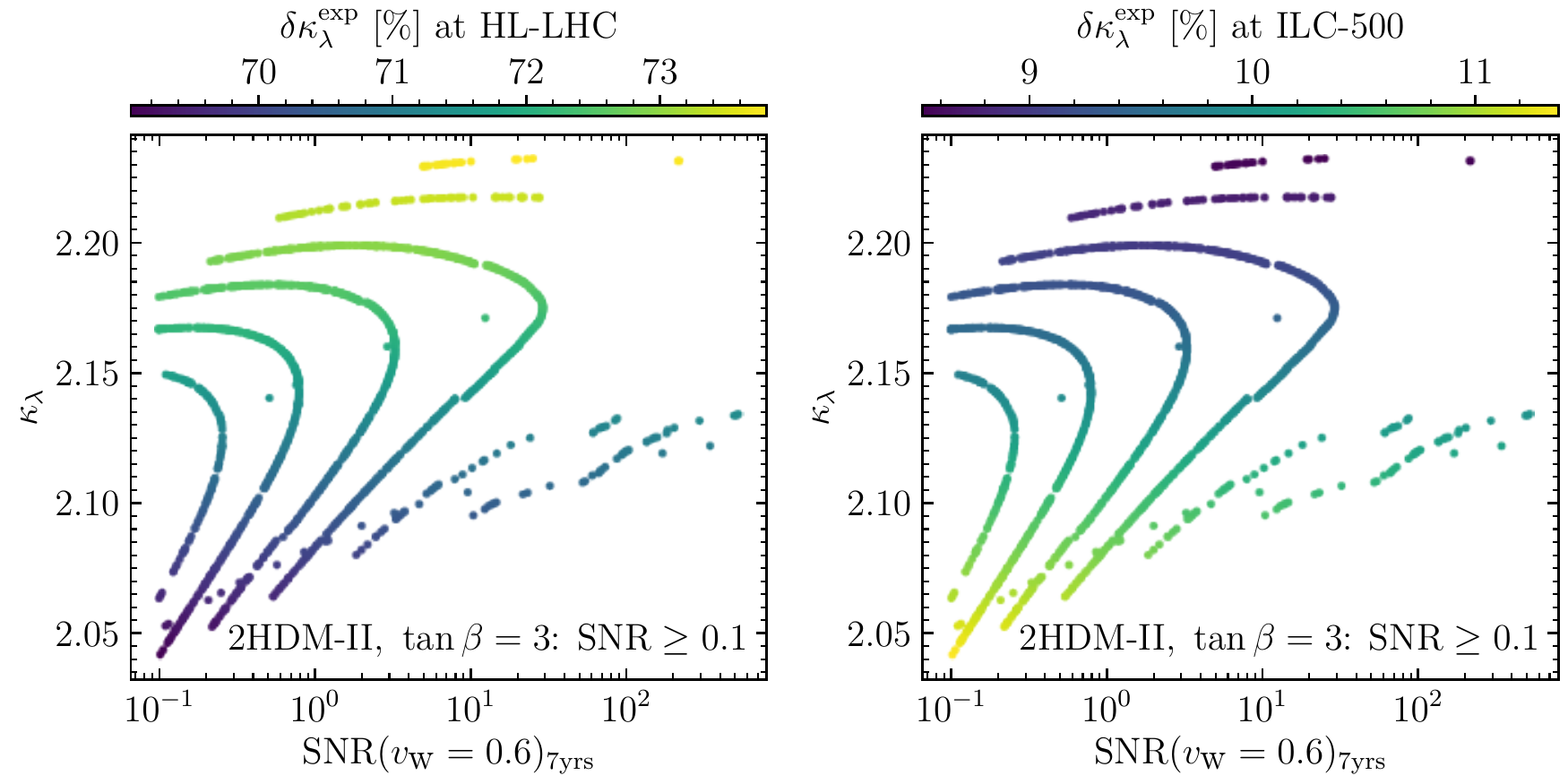}
\caption{\small
  Parameter points from \reffi{fig7b} with $\textrm{SNR} \geq 0.1$ in the (SNR, $\kappa_\lambda$) plane. The color coding of the points indicates the projected experimental precision of the measurement of $\kappa_\lambda$ at
  the HL-LHC (left) and the ILC-500 (right), see text for details.}
\label{fig:cplsprec}
\end{figure}

\vspace{1mm}

In all the above discussion, we have focused on the potential of HL-LHC and ILC 
\GW{constraints on} $\kappa_\lambda$ to exclude the presence of sizable BSM contributions to $\lambda_{hhh}$ by assuming 
\GW{that the value of}
$\kappa_\lambda = 1$ 
\GW{is realized in nature. However, as explained above the parameter region of the 2HDM giving rise to a FOEWPT predicts values of $\kappa_\lambda$ that are significantly larger than the SM value. Therefore it is also important to assess the capabilities of the HL-LHC and the ILC (or another $e^+e^-$ Linear Collider running at $500\gev$) for making a measurement of the trilinear Higgs boson coupling for the case where $\kappa_\lambda \neq 1$. 
In fact,} the expected HL-LHC and ILC precision of the $\kappa_\lambda$ measurement, $\delta \kappa_{\lambda}^{\rm exp}$, would significantly change 
\GW{for the case where the true value of $\kappa_\lambda$ is above 1}
(for $\kappa_\lambda = 1$ the \GW{projected} HL-LHC and ILC-500 precisions are given by $\delta \kappa_{\lambda}^{\rm exp} = 60\%$~\cite{Cepeda:2019klc} and $\delta \kappa_{\lambda}^{\rm exp} = 27\%$~\cite{Durig:2016jrs}, respectively).
In order to analyze how precisely the HL-LHC and the ILC could measure a value of $\kappa_\lambda$ in the 2HDM parameter space region yielding an observable GW signal at LISA, we show in \reffi{fig:cplsprec} the parameter points of \reffi{fig7b} with $\textrm{SNR} \geq 0.1$ in the (SNR, $\kappa_\lambda$) plane, with the color-coding indicating the experimental precision with which $\kappa_{\lambda}$ could be measured at the HL-LHC (left panel) and the ILC-500 (right panel). At the HL-LHC, the experimental precision of a $\kappa_\lambda \sim 2$ measurement  ($\delta \kappa_{\lambda}^{\rm exp} \gtrsim 70\%$) worsens compared to that of $\kappa_\lambda = 1$. This is due to the enhanced 
\GW{destructive}
interference between 
\GW{the contributions involving and not involving the trilinear Higgs coupling}, leading to a reduced cross section at the HL-LHC (see, for instance, Fig.~3 of \citere{Frederix:2014hta} for the cross-section predictions).
On the other hand, the situation would be much more favorable at the ILC with $\sqrt{s} = 500 \gev$ in the process $e^+e^- \to Zhh$, for which a precision of $\delta \kappa_{\lambda}^{\rm exp} \sim 10\%$ could be expected for a $\kappa_\lambda \sim 2$ measurement with an integrated luminosity of $4000\, \mathrm{fb}^{-1}$~\cite{Durig:2016jrs}.  
The Higgs boson self-coupling measurement at the ILC-500 relies mainly on the Higgs-strahlung channel, which exhibits a 
\GW{constructive} interference between 
\GW{the different contributions} and thus an enhanced di-Higgs production cross section for $\kappa_\lambda > 1$ (see Fig.~8 of \citere{LCCPhysicsWorkingGroup:2019fvj}).

\vspace{1mm}

\GW{Regarding} the interpretation of \reffi{fig:cplsprec} we \GW{would like to} remind the reader that the current theoretical uncertainties on the prediction for the GW spectra from a FOEWPT, as well as the lack of knowledge of the value of $v_{\rm w}$ (see section~\ref{sec:gws} for details), translate into an uncertainty on the SNR (not shown in the plots) 
\GW{that is much
larger than the one that is induced by a collider measurement of the trilinear Higgs coupling with an uncertainty of $\delta \kappa_{\lambda}^{\rm exp} \sim 10\%$ (reachable at the ILC-500).
Thus, a measurement of $\kappa_\lambda \sim 2$, 
which would be possible only with rather large uncertainty at the HL-LHC but with a much better precision at the ILC-500 or other $e^+e^-$ colliders running at a similar energy, together with a non-observation of a GW signal at LISA
clearly would not rule out a 2HDM
\htm{(type~II)} interpretation.
On the other hand, if a value of
$\kappa_\lambda \sim 1$ is established via collider measurements and a GW signal at LISA is detected, an interpretation within the 2HDM would be strongly disfavored.}

\section{Conclusions}
\label{conclu}

In this paper we have analyzed the thermal history of the 2HDM, focusing on its type~II variant, and its associated phenomenological imprints. 
It is well known that the 2HDM (in contrast to the SM) can accommodate a strong first-order electroweak phase transition (FOEWPT). \GW{The FOEWPT can} lead to the production of a primordial gravitational wave (GW) background \GW{that could} potentially be detectable by the future LISA observatory, and/or allow for the required out-of-equilibrium conditions in order to realize successful baryogenesis.
We have shown that the 2HDM may also give rise to other phenomena during its thermal evolution in the early Universe, characterized by vacuum trapping (the Universe remains in an unbroken EW phase, although the EW vacuum is the deepest one at $T=0$) and EW symmetry non-restoration (SnR), the possibility that the vacuum adopted at high temperature is not the EW symmetric one. 
Within a simple scenario characterized by the alignment limit ($c_{\beta-\alpha} = 0$) and equal masses for the neutral CP-odd and charged BSM scalars ($m_A = m_{H^\pm}$), we have categorized the different thermal histories which are possible in the $(m_H,m_A)$ plane of the 2HDM:
A)~the Universe always stays in the EW vacuum (SnR), although at $T=0$ the EW broken phase is meta-stable; B)~the Universe always remains at the EW symmetric state even though a meta-stable (with a lifetime longer than the age of the Universe) EW broken minimum is present at $T=0$; C)~the Universe always remains in the EW broken phase (SnR), which is always the deepest minimum of the potential; D)~the Universe always remains in the EW symmetric phase, although at $T=0$ the EW vacuum
is the deepest one (vacuum trapping); E)~the Universe undergoes a strong FOEWPT (from the EW symmetric to the broken phase); F)~the Universe undergoes a weak first-order or a second-order transition (from the EW symmetric to the broken phase). The fact that regions~B and~D are unphysical allowed us to determine new limits on the 2HDM parameter space.
In particular, regarding the occurrence of vacuum trapping (region D), we have demonstrated that it excludes a sizable region of parameter space that would otherwise feature a strong FOEWPT, stressing the importance of the determination of the false vacuum decay rate and the nucleation temperature at which the EW phase transition does take place.
Merely relying on the presence of a critical temperature at which the co-existing EW symmetric and EW broken vacua are equally deep (as frequently applied in the literature, not accounting for vacuum trapping) erroneously assigns the strongest FOEWPTs to regions of the 2HDM parameter space in which actually no transition can occur.

\vspace{1mm}

Focusing our analysis on the (type~II) 2HDM parameter region featuring a strong FOEWPT, we have found that, even with most optimistic assumptions (bubble-wall velocity $v_{\rm w} = 0.6$, 
\GW{taking into account the turbulent motion of the primordial plasma as source for GWs,} and $\mathcal{T} = 7$ years of effective LISA observation time), GW signals from the EW epoch \GW{that are} potentially observable by LISA (with a signal-to-noise ratio ${\rm SNR} > 1$) only occur in a very 
\GW{narrow}
region of the 
\GW{$(m_H, m_A)$ mass plane of the 2HDM}, 
corresponding to very specific values of the mass-splitting $m_A - m_H$ (which are generally large, $200\gev \lesssim m_A - m_H \lesssim 250\gev$) as a function of $m_H$. 
Parameter regions with larger mass splittings either feature SnR (and so do not give rise to a FOEWPT) or are unphysical. 
\GW{Indeed, we found that the parameter region giving rise to the strongest GW signals is adjacent to the (unphysical) parameter region featuring vacuum trapping,}
\htm{and we demonstrated that this fact gives
rise to the very strong dependence of the
amplitude and peak frequency of the
potentially detectable GW signals on the underlying
model parameters.}

In addition, we have explored the collider phenomenology of 2HDM parameter regions yielding a strong FOEWPT, including those generating a GW signal 
\GW{that could be} within the future reach of LISA, in order to assess the interplay between the LHC, LISA and future colliders like the ILC to scrutinize FOEWPT scenarios in the 2HDM.
First, based on the RGE evolution of 2HDM quartic scalar couplings, the existence of new strongly-coupled physics (beyond the 2HDM) would be needed at energy scales $\Lambda_{4\pi} \sim 1 - 2$ TeV for scenarios yielding an observable GW signal at LISA. We can thus safely argue that such scenarios should be within reach of the (HL-)LHC. At the same time, we have demonstrated that the 2HDM parameter regions that LISA could probe would yield an LHC 
``smoking gun signature'' $pp \to A \to Z\,H, \, \, H \to t \bar t$. A conservative extrapolation of the public preliminary results in this channel by the CMS collaboration to HL-LHC integrated luminosities shows that this search would cover \GW{essentially} 
the entire region \GW{that could be} observable by LISA. Again, this 
\GW{has a crucial impact on} 
the possible interplay between LISA and the LHC:
the absence of new-physics indications at the (HL-)LHC would make the observation of a GW signal from a FOEWPT in the 2HDM by LISA virtually impossible.

\GW{As a final step of our analysis we focussed on the trilinear self-coupling of the Higgs boson at about $125\gev$, $\lambda_{hhh}$. We pointed out that
the measurement of this coupling} constitutes an important probe of a FOEWPT in the early Universe, since FOEWPT scenarios are generically associated with sizable enhancements of $\lambda_{hhh}$ with respect to its SM value. We 
\GW{have found} that 
regions in the 2HDM parameter space that give rise to GW signals with sizable signal-to-noise ratios at LISA (${\rm SNR} > 0.1$) are associated with large values of $\kappa_\lambda = \lambda_{hhh}/\lambda_{hhh}^{\rm SM} \sim 2$, and even larger values of $\kappa_\lambda$ are found for SnR scenarios. 
If $\kappa_\lambda \sim 1$ (as in the SM) is realized in nature, $\kappa_\lambda \sim 2$ values are at the border of the 95\% C.L.\ upper limits expected from the measurement of the non-resonant Higgs-boson pair-production at the HL-LHC. Then, SnR scenarios in the 2HDM will be well-probed by this measurement, which will also access the 2HDM parameter region yielding the strongest GW signals. 
\GW{If on the other hand indeed a value of $\kappa_\lambda \sim 2$ is realized in nature, it is important to note}
that 
the precision with which $\kappa_\lambda$ can be measured significantly depends on its precise value. \GW{A value of} $\kappa_\lambda \sim 2$ leads to a reduced sensitivity at the HL-LHC, with a precision of only $\sim 70\%$ (compared to $\sim 60\%$ for $\kappa_\lambda \sim 1$) due to the enhanced \GW{destructive} interference of 
\GW{contributions involving and not involving the trilinear Higgs coupling}, leading to a reduced 
Higgs boson pair production cross section. The situation is reversed at the ILC operating at $\sqrt{s} = 500 \gev$: for $\kappa_\lambda = 1$ a~\GW{$\sim 27\%$} precision on the measurement is anticipated, while for $\kappa_\lambda \sim 2$ the precision increases to~$\sim 10\%$ due to an enhanced 
\GW{constructive interference between the different contributions}. Since a FOEWPT is naturally connected to \GW{values of} $\kappa_\lambda > 1$, the general prospects \GW{in this case} for the HL-LHC to measure the Higgs boson self-coupling are worse than in the SM, whereas they improve substantially for the ILC.

\GW{Accordingly, a collider measurement of $\kappa_\lambda \sim 2$, 
would be well compatible with a 2HDM (type~II) interpretation, independently of whether or not a GW signal at LISA will be detected. On the other hand, 
a collider measurement of
$\kappa_\lambda \sim 1$ together with a GW signal at LISA would strongly disfavor an interpretation within the type~II 2HDM.}

\section*{Acknowledgements}
We thank Daniel Hundhausen and Matthias Schr\"oder \htk{from the CMS Group at the University
of Hamburg} for interesting discussions. \htk{We are also grateful to Jenny List for the projections of the HL-LHC and ILC sensitivities to the triple Higgs coupling measurements.} T.B., M.O.O.R.~and G.W.~acknowledge support by the Deutsche Forschungsgemeinschaft (DFG, German Research Foundation) under Germany‘s Excellence Strategy -- EXC 2121 ``Quantum Universe'' – 390833306. This work has been partially funded by the Deutsche Forschungsgemeinschaft 
(DFG, German Research Foundation) - 491245950. 
\htk{S.H and J.M.N. acknowledge support from the grant IFT Centro de Excelencia Severo Ochoa
CEX2020-001007-S funded by MCIN/AEI/10.13039/501100011033.} 
The work of S.H.\ was supported in part by the
grant PID2019-110058GB-C21 funded by
MCIN/AEI/10.13039/501100011033 and by ``ERDF A way of making Europe''.
The work of J.M.N. was supported by the Ram\'on y Cajal Fellowship contract RYC-2017-22986, and by grant PGC2018-096646-A-I00 from the Spanish Proyectos de I+D de Generaci\'on de Conocimiento.
J.M.N. also acknowledges support from the European Union's Horizon 2020 research and innovation programme under the Marie Sklodowska-Curie grant agreement 860881 (ITN HIDDeN).

\appendix

\section{Comparison of the renormalization
scale dependence with changes of the 2HDM parameters}
\label{app:scale}

In \refse{sec: renorm2hdm} we discussed the impact
of the renormalization scale dependence
of the effective potential on
the predicted values for the quantities
that characterize the FOEWPT.
We 
pointed out that the $\mu$-dependence
is much smaller
compared to the large dependence on the
values of the 
model parameters.
In \refse{sec:gravitationalwaves}
it was shown that the dependence on the model parameters is particularly
large in the parameter space regions that
could potentially feature a FOEWPT that is
sufficiently strong 
to give rise to
a detectable GW signal.
As a consequence, different choices for
the renormalization scale $\mu$
(within a physically reasonable range) would
not have a major impact on the
prospects 
for a detectable GW
signal in certain regions of the
2HDM parameter space.

\begin{figure}[t]
\centering
\includegraphics[width=0.6\textwidth]{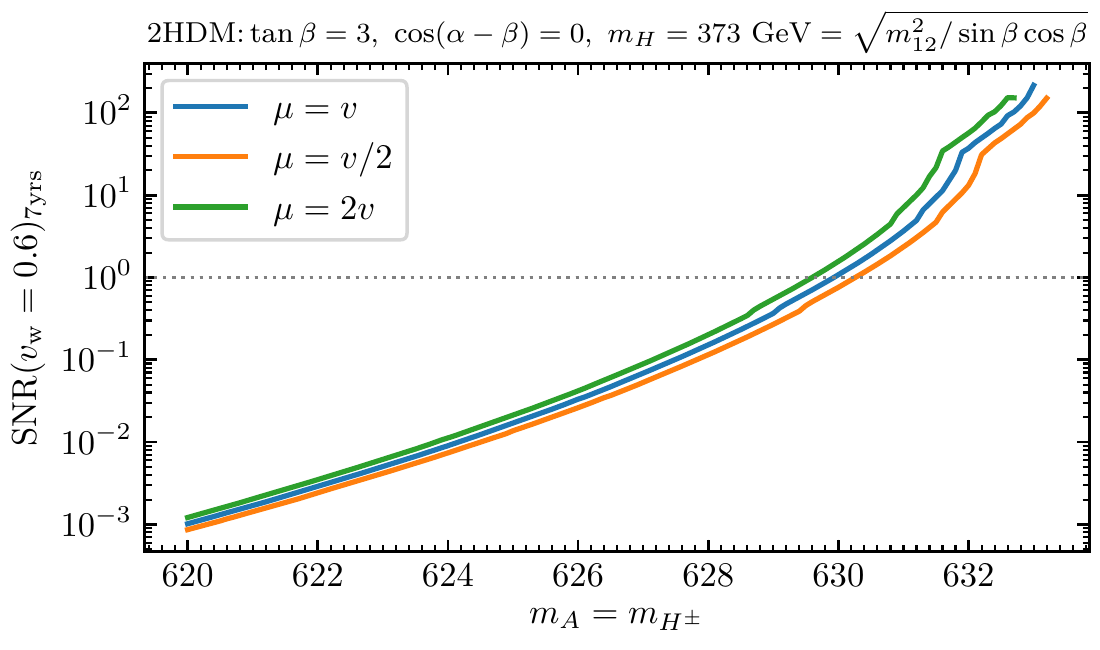}
\caption{\small 
LISA SNR of the GW signal as
a function of $m_A = m_{H^\pm}$
(all other parameters are kept fixed)
for an example scenario of the scan
discussed in \refse{sec:2hdm_thhist} for three
different choices of the renormalization
scale~$\mu$.}
\label{fig:scaledep}
\end{figure}

In order to illustrate this, we show in \reffi{fig:scaledep}
the predictions for the LISA SNR of the GW signals
in an example scenario taken from the scan
discussed in \refse{sec:2hdm_thhist}.
The 2HDM parameters are set as shown in
\refeq{eqranges1}, but for a fixed value of $m_H = 373\gev$,
and the masses $m_A = m_{H^\pm}$ are varied in
the small range in which potentially observable
GW signals are predicted.
We depict the SNR as a function of 
$m_A = m_{H^\pm}$ for three different values of
the renormalization scale: $\mu = v, v/2$ and
$2v$ indicated in blue, orange and green, respectively.
One can make several observations that 
demonstrate
the robustness of our conclusions 
with respect to the
theoretical uncertainties stemming from the
renormalization scale dependence:

\textit{(i)} Apart from a very small range of
mass values, the predicted SNRs of all three
curves are orders of magnitudes below 
$1$,
indicating that 
those predicted GW signals would not
be observable at LISA, independently of the
choice for the renormalization scale $\mu$.
Therefore, our conclusions that potentially detectable
GW signals only occur in very narrow regions
of the analyzed $(m_H,m_A)$ mass plane do not
depend on the precise value of $\mu$.

\textit{(ii)} For the mass ranges in which the
predicted SNRs are of the order of 
$1$ or larger,
indicating that the corresponding parameter
space regions could potentially be probed with LISA,
we find that the variation of $\mu$ gives rise
to variations of the SNRs by about an order of
magnitude. However, the same variation of the
SNRs 
occurs
(for a fixed value of $\mu$)
if $m_A = m_{H^\pm}$ is varied by less than
$1\gev$. This can be seen, for example, by comparing
the values of $m_A = m_{H^\pm}$ at which the
three colored lines cross the horizontal
dashed gray line at $\mathrm{SNR} = 1$.
This value changes from $m_A = m_{H^\pm} = 629.5 \gev$ 
for $\mu = v/2$ to $m_A = m_{H^\pm} = 630.0 \gev$ for 
$\mu = 2 v$.
As a consequence, the 
minimum amount of the mass splittings
between $H$ and $A,H^\pm$ that is required 
for a sufficiently strong GW signal is only
marginally affected by the variation of $\mu$.

\textit{(iii)} The three colored lines 
end at a maximum value of $m_A = m_{H^\pm}$ 
as a consequence
of vacuum trapping (see the discussion
in \refse{sec:vactrapp}). The 
endpoints of the three lines
lie within
a range of less then 1~GeV
in $m_A = m_{H^\pm}$, such that also the maximal
mass splitting between $H$ and $A,H^\pm$ that can
be realized without the occurence of vacuum trapping
is not significantly affected by the variation of $\mu$.

In summary, our conclusions about the interplay between collider physics at the
\mbox{(HL-)LHC}
and the possible observation of GW signals at LISA
in the context of the type~II 2HDM
are robust against the uncertainties from the
renormalization scale dependence of the
effective potential.

\bibliographystyle{JHEP}
\bibliography{lit}

\end{document}